\documentclass{aa}
\usepackage{psfig}

\def\h2{H{\small II}}
\newcounter{qub}
\def\sbs{SBS 0940+544}

\begin{document}

\title{The evolutionary status of the low-metallicity blue compact 
dwarf galaxy SBS 0940+544}

%
%

\author{N. G. Guseva \inst{1}
\and Y. I. Izotov \inst{1}
\and P. Papaderos \inst{2}
\and F. H. Chaffee \inst{3}
\and C. B. Foltz \inst{4}
\and R. F. Green \inst{5}
\and T. X. Thuan\inst{6}
\and K. J. Fricke \inst{2}
\and K. G. Noeske\inst{2}}
\offprints{guseva@mao.kiev.ua}
\institute{      Main Astronomical Observatory,
                 Ukrainian National Academy of Sciences,
                 Golosiiv, Kyiv 03680,  Ukraine
\and
                 Universit\"ats--Sternwarte, Geismarlandstra\ss e 11,
                 D--37083 G\"ottingen, Germany
\and
                 W. M. Keck Observatory, 65-1120 Mamalahoa Hwy., Kamuela,
                 HI 96743, USA
\and
                 MMT Observatory, 
                 University of Arizona, 
                 Tucson, AZ 85721, USA
\and
                 National Optical Astronomy Observatories, 
                 Tucson, AZ 85726, USA
\and
                 Astronomy Department, University of Virginia, 
                 Charlottesville, VA 22903, USA
}

\date{Received \hskip 2cm; Accepted}

\abstract{We present the results of spectrophotometry and $V, R, I$, H$\alpha$ 
CCD photometry of the blue compact dwarf (BCD) galaxy SBS~0940+544.
Broad-band images taken with the
2.1m KPNO\thanks{Kitt Peak National Observatory (KPNO) 
is operated by the Association of Universities for Research in Astronomy, 
Inc., under cooperative agreement with the National Science Foundation.} 
and 1.23m Calar Alto\thanks{German--Spanish Astronomical Center, 
Calar Alto, operated by the Max--Planck--Institute for Astronomy, Heidelberg, 
jointly with the Spanish National Commission for Astronomy.} telescopes
reveal a compact high-surface-brightness H {\sc ii} region 
with ongoing star formation, located at the northwestern tip
of the elongated low surface brightness (LSB) 
main body of the BCD.
Very faint, patchy emission along the main body is seen in the H$\alpha$ image.
High signal-to-noise 4.5m Multiple Mirror Telescope
(MMT)\thanks{The MMT Observatory is a joint facility of the 
Smithsonian Institution and the University of Arizona.} 
and 10m Keck II telescope\thanks{W.M. Keck Observatory 
is operated as a scientific partnership among the California Institute of 
Technology, the University of California and the National Aeronautics and Space 
Administration. The Observatory was made possible by the generous financial 
support of the W.M. Keck Foundation.} long-slit spectroscopy of 
SBS~0940+544 is used to derive element abundances of the ionized gas in the 
brightest H {\sc ii} region and to study the stellar
population in the host galaxy. The oxygen abundance in the brightest region 
with strong emission lines is 12 + log(O/H) = 7.46~--~7.50, or 
1/29~--~1/26 solar, in agreement with earlier determinations and among
the lowest for BCDs.
H$\beta$ and H$\alpha$ emission lines and H$\delta$ and H$\gamma$ absorption
lines are detected in a large part of the main body. 
Three methods are used to put constraints on the age of the stellar 
population
at different positions along the major axis.
They are based on (a) the equivalent widths of the emission lines, 
(b) the equivalent widths of the 
absorption lines and (c) the spectral energy distributions (SED). 
Several scenarios of star formation have been considered. 
We find that models with single instantaneous bursts
cannot reproduce the observed SEDs implying that star formation 
in the main body of SBS 0940+544 was continuous.
%
The observed properties in the main body can be reproduced  
by a continuous star formation process which started not earlier than
100 Myr ago, if a small extinction is assumed. However,
the observations can be reproduced equally well 
by a stellar population forming continuously since 10 Gyr ago, if 
 the star formation rate has increased during the last 100 Myr in the main 
body of SBS 0940+544 by at least a factor of five.
%
We also constrain the age of the reddest southern region of the main body,
where no absorption and emission lines are detected. 
On the assumption of zero extinction in this faint region, the observed 
spectrum can be fitted by a theoretical SED of a stellar population 
continuously formed with constant star formation rate between 100 Myr and
10 Gyr ago. If, however, a small extinction of $C$(H$\beta$) $\sim$ 0.1 is 
present in this region then the observed spectrum can be fitted by a 
theoretical SED of a stellar population continuously formed  with a constant 
star formation rate between 100 Myr and 1 Gyr ago. However, the poor 
signal-to-noise ratio of the spectrum and large photometric errors preclude 
reliable determination of the age of the southern region. 
In summary, we find no compelling evidence which favors either a young or an 
old age of SBS 0940+544.
%
}

\maketitle

\markboth {N. G. Guseva et al.}{The evolutionary status of the 
BCD SBS 0940+544}

\section{Introduction}

Blue compact dwarf (BCD) galaxies are characterized by ongoing intense 
star formation (SF) activity as evidenced by strong nebular emission 
lines superposed on a blue continuum
(e.g. Sargent \& Searle \cite{SS70}; Lequeux et al. \cite{L79};
Kunth \& Sargent \cite{K83}; Izotov, Thuan \& Lipovetsky \cite{ITL94}).
They are compact and metal-deficient
objects with oxygen abundances ranging between 1/50 and 1/3 solar 
(e.g. Izotov \& Thuan \cite{IT99}) and possess large amounts of neutral gas
(Thuan \& Martin \cite{TM81}; van Zee et al. \cite{VanZee98};
Thuan et al. \cite{TLMP99}). 

Most of the BCDs studied so far have revealed extended low surface brightness 
red stellar sheets underlying the blue star-forming regions 
(Thuan \cite{Thuan83}; Loose \& Thuan \cite{Loose86}). 
Some of these  stellar sheets have been resolved with the Hubble Space 
Telescope (HST) into red giant stars (Schulte-Ladbeck, Crone \& Hopp 
\cite{SCH98}; Crone et al. \cite{Crone2000}). 
This indicates the presence of an evolved low-mass stellar populations
in the underlying host galaxies.

The tiny fraction of galaxies with extremely low oxygen abundance
in the ionized gas (less than  1/20 (O/H)$_{\odot}$) has been suggested
by Izotov \& Thuan (\cite{IT99}) to be young galaxies, based on chemical
element abundance arguments. 
Nearly ten such galaxies with good abundance determinations are known to 
date. The outer parts of 
these galaxies are blue. The spectral energy distributions and 
colours of their outskirts are consistent with those of stellar 
populations with ages less than a few 100 Myr (see the examples
of SBS 0335--052 (Izotov et al. \cite{ILCFGK97c}, Thuan, Izotov \& 
Lipovetsky \cite{til97}, Papaderos et al. \cite{Papa98}) and SBS 1415+437 
(Thuan, Izotov \& Foltz \cite{TIF99})).

 Detailed spectroscopic and photometric studies of extremely metal-poor BCDs
based on observations with large telescopes are useful to shed light on the 
origin of these galaxies, deduce their star formation history and constrain
their ages. 

A good candidate for being a young galaxy is SBS 0940+544.
This galaxy was discovered in the course of the Second Byurakan Survey (SBS, 
Markarian, Lipovetsky \& Stepanian \cite{Mark83}) and described as a compact 
extragalactic H {\sc ii} region without an underlying host galaxy, with
strong emission lines, a weak continuum and an ultraviolet excess.
 
A first abundance determination in the brightest H {\sc ii} region of
SBS 0940+544 was made by Izotov et al. (\cite{I91})
who derived an oxygen abundance 12 + log(O/H) = 7.52 $\pm$ 0.12 and a
helium mass fraction $Y$ = 0.246 $\pm$ 0.030. Izotov, Thuan \& Lipovetsky
(\cite{Izotov97}), using higher S/N ratio observations, derived 
subsequently a lower value 12 + log(O/H) = 7.43 $\pm$ 0.01. 
Because of its very low metal content, SBS 0940+544 was used for the
determination of the primordial helium abundance (Izotov, Thuan 
\& Lipovetsky \cite{ITL94}; Izotov \& Thuan \cite{IT98a}) and to study
heavy element abundances (Thuan, Izotov \& Lipovetsky \cite{til95};
Izotov \& Thuan \cite{IT99}).  
The galaxy has been studied spectroscopically also by 
Augarde et al. (\cite{Augarde94}) and Comte et al. (\cite{Comte94}), without
an element abundance determination.

$B$ and $R$ surface photometry and colour profiles of \sbs\ have been derived 
by Doublier et al. (\cite{Doublier97}) (see also 
Doublier, Caulet \& Comte (\cite{Doublier99})).

Correcting the 21 cm heliocentric velocity $V$ = 1638 km s$^{-1}$
for Virgocentric infall with a velocity of 220 km s$^{-1}$ and adopting 
a Hubble constant of 75 km s$^{-1}$ Mpc$^{-1}$, Thuan et al. 
(\cite{TLMP99}) derived a distance of $D$ = 26.7 Mpc for SBS 0940+544.
At this distance, assumed here throughout, 1\arcsec\ corresponds to 
a projected linear scale of 129 pc.
The total neutral hydrogen H {\sc i} mass in SBS 0940+544 is 
log($M$(H {\sc i})/$M_\odot$) = 8.71 
(or $M$(H {\sc i}) =5.1 $\times$ $10^8$ $M_\odot$)
(Thuan et al. \cite{TLMP99}).

In this paper we present new $V,R,I$ and H$\alpha$ photometry and
very high signal-to-noise spectroscopic observations of
SBS 0940+544. We use those data to study the stellar populations in the BCD,
put constraints on its age and derive elemental
abundances in the ionized gas, including the helium abundance.
In Sect. 2 we describe the observations and data reduction. 
The photometric properties of SBS 0940+544 are
discussed in Sect. 3. We derive elemental abundances in the brightest H {\sc ii}
region in Sect. 4. In Sect. 5 the stellar populations are
discussed and constraints on the age of SBS 0940+544 are considered. 
Sect. 6 summarizes the results.

\section{Observations and data reduction}

\subsection{Photometry}

Direct images of SBS 0940+544 were acquired in two observing runs. Broad-band 
Johnson $V$ and Cousins $I$ images and narrow-band images in the 
H$\alpha$ line at $\lambda$6606\AA\ with a full width at half maximum
(FWHM) through a passband of 78\AA\ and 
in the adjacent continuum at $\lambda$6520\AA\ through a passband with 
FWHM = 84\AA\ were obtained with the Kitt Peak 2.1m telescope 
on April 18 and April 22, 1999, respectively. The telescope was equipped 
with a 1k 
Tektronix CCD detector operating at a gain of 3\,e$^-$\,ADU$^{-1}$, giving an 
instrumental scale of 0\farcs 305 pixel$^{-1}$ and field of view of 5\arcmin. 
The total exposure of 40 and 60 min in $V$ and $I$, respectively, 
was split up into four subexposures, each being slightly offset from the
others. Likewise, narrow-band exposures in the H$\alpha$ line and the  
adjacent continuum bluewards of H$\alpha$ were taken for 25 min each. 
The seeing during the 
broad- and narrow-band observations was FWHM 1\farcs 2 and 2\farcs 2, 
respectively.

Another series of exposures with a total integration time 
of 2.5 hours in Cousins $R$ was taken with the 1.23m telescope 
at Calar Alto
in the period January 24 to February 22, 2000. The telescope was equipped
with a 2k 
SITe detector at its Cassegrain focus giving a usable 
field of view of $\sim$ 11\arcmin\
and an instrumental scale of 0\farcs5 pixel$^{-1}$. The FWHM of 
point sources in the coadded $R$ exposures is $\approx$ 1\farcs 5.

 Bias and flat--field frames were obtained during each night of 
both observing runs. Broad-band images were calibrated by observing 
different standard fields from Landolt (\cite{Landolt92}). 
Our calibration uncertainties are estimated to be well below 0.05\,mag in all 
bands. Standard reduction steps, including bias subtraction, flat--field 
correction and absolute flux calibration were carried out using the 
IRAF\footnote{IRAF is the Image 
Reduction and Analysis Facility distributed by the 
National Optical Astronomy Observatory, which is operated by the 
Association of Universities for Research in Astronomy (AURA) under 
cooperative agreement with the National Science Foundation (NSF).}
and ESO 
MIDAS\footnote{Munich Image Data Analysis System, provided by the European 
Southern Observatory (ESO).}.

\subsection{Spectroscopy \label{obsspec}}

High S/N long-slit spectroscopy of SBS~0940+544 was obtained  
with the 4.5m Multiple Mirror Telescope  (MMT)
and with the 10m Keck II telescope over a wavelength range 3600 -- 7500\AA.

\begin{figure*}[!ht]
    \hspace*{0.9cm}\psfig{figure=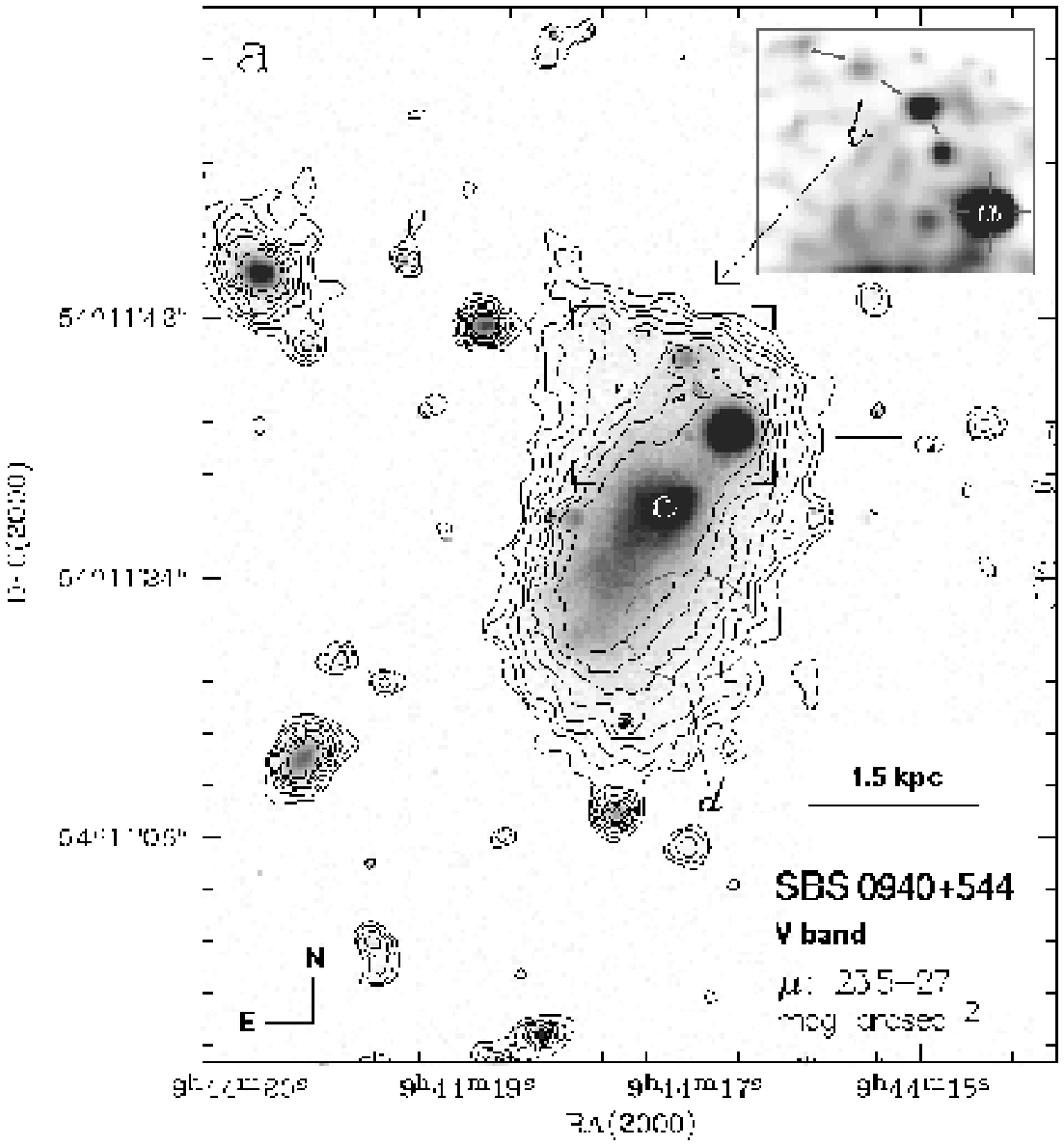,angle=0,height=10cm,clip=}
    \hspace*{1.7cm}\psfig{figure=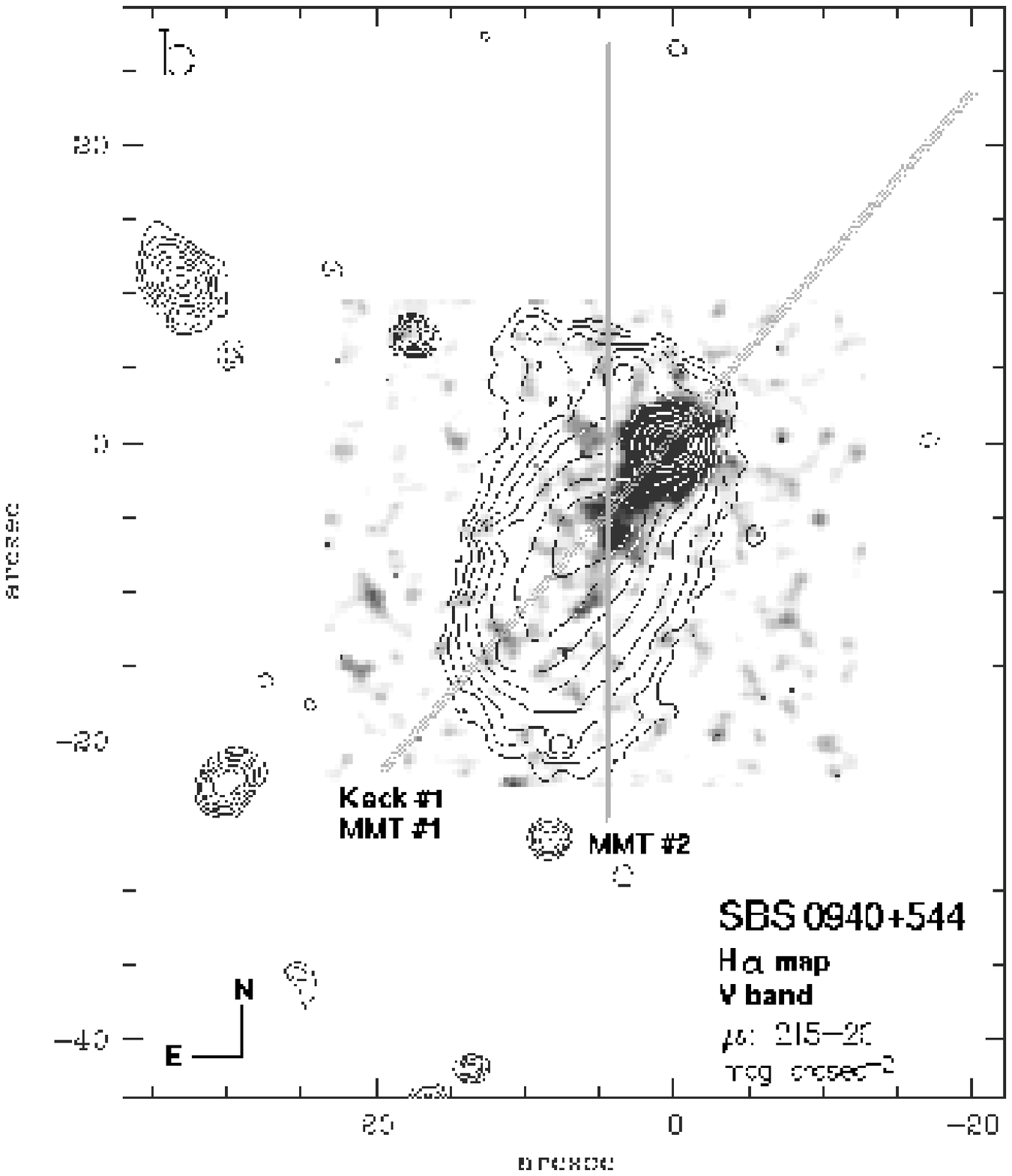,angle=0,height=10cm,clip=}
   \caption[]{(a) $V$ image of the {\it cometary} iI BCD 
SBS 0940+544. The overlayed contours correspond to surface brightness levels 
between 23.5 mag arcsec$^{-2}$ and 27 mag arcsec$^{-2}$ in steps of 0.5 mag. 
The inset at the upper right shows a deconvolved 
13\arcsec\ $\times$ 11\arcsec\ subframe at 
the northwestern tip of the galaxy. Adjacent to the brightest H {\sc ii} region
{\it a} is a chain of other compact sources (labeled {\it b}) traceable to 
$\sim$ 10\arcsec\ northeast of {\it a}. From the continuum-subtracted 
H$\alpha$ map of SBS 0940+544 (right panel) and from spectroscopic study 
(Sect. 4), star-formation activity is found to be also 
present close to the optical maximum of the LSB body of the BCD (labeled 
{\it c}). 
The average $V-R$ and $V-I$ colours of the main body increase from 
$\sim$0.16 mag and $\sim$0.43 mag in region {\it c} to $\sim$0.34
mag and $\sim$0.65 mag in region {\it d}. 
(b) Continuum-subtracted H$\alpha$ map of SBS 0940+544 overlayed with $V$ band 
contours
between 21.5 mag arcsec$^{-2}$ and 26 mag arcsec$^{-2}$ in steps of 0.5 mag.
Roughly 80\% of the H$\alpha$ emission originates from the compact 
H {\sc ii} region {\it a}. 
The long-slit positions during the two observations with the 
Multiple Mirror Telescope 
(MMT \#1 and MMT \#2) and with the Keck\,II telescope (Keck \#1) are indicated.}
  \label{f1}
  \end{figure*}

 The first set of observations was carried out with the MMT
at two slit positions (see Fig.~\ref{f1}b). Observations were made with the 
blue channel of the MMT spectrograph operated with a
3072~$\times$~1024 CCD detector. A 1\farcs5~$\times$~180\arcsec\ slit was 
used. The spatial scale along the slit was 0\farcs3~pixel$^{-1}$, with a
spectral resolution of $\sim$ 7~\AA\ (FWHM). The final spatial scale
is 0\farcs6~pixel$^{-1}$ after binning every two consecutive rows into one.

 The first slit orientation with position angle P.A.=--41$^{\circ}$
(MMT \#1, cf. Fig. 1b), was 
centered on the brightest star-forming H {\sc ii} region 
(denoted as {\it a} in Fig.1a) and oriented along the 
elongated main body of the galaxy close to the direction of the major axis.
This 120-minute spectrum was obtained on December 13, 1996
at an airmass of 1.2. The total exposure time was broken into four 30 min 
exposures.
 
 The second slit orientation (MMT \#2, cf. Fig. 1b)
is in the N-S direction with P.A. = 0$^{\circ}$, 
centered on the second brightest region, denoted as {\it c} in 
Fig.~\ref{f1}a. These observations were carried out on December 12, 
1996 with an exposure time of 120 minutes at an airmass of 1.2.
The total exposure time was also broken into four 30 min 
exposures. The seeing during both sets of MMT observations was 0\farcs8.

 The Keck II observations were carried out
on January 9, 2000 with  the low-resolution imaging spectrograph (LRIS)
(Oke et al. \cite{Oke95}), using the
300 groove mm$^{-1}$ grating which provides a dispersion 
2.52 \AA\ pixel$^{-1}$ and a
spectral resolution of about 8 \AA\ in first order. 
The slit was 
1\arcsec$\times$180\arcsec, centered on the brightest H {\sc ii} region 
and oriented with 
a position angle P.A. = --41$^{\circ}$, the same as during the first set of MMT
observations. No binning along the spatial axis has been done, yielding a 
spatial sampling of 0\farcs2 pixel$^{-1}$. The total exposure time was 60 min,
broken into three
20 min exposures. All exposures were taken at an airmass of
1.2. The seeing was 0\farcs9.

Signal-to-noise ratios 
S/N $\sim$ 50 and $\sim$ 100 were reached in the continuum near H$\beta$ 
for the brightest part of the galaxy during the MMT and Keck II observations.
No correction for atmospheric dispersion was made because of 
the small airmass during all observations. 
This effect, although small, may introduce some uncertainties in the flux
determination of the weaker lines.
At an airmass of 1.2 and position angle
--41$^{\circ}$ the shift of the emitting region at the [O {\sc ii}] 
$\lambda$3727 line relative to the same region at the 
[O {\sc iii}] $\lambda$5007 
line in the direction perpendicular to the slit is 0\farcs4 (Filippenko 
\cite{F82}).
The shift at [O {\sc iii}] $\lambda$4363 is much smaller, only 0\farcs1. 
In Sect. \ref{chem} we estimate the uncertainties introduced by this effect 
on the abundance determination. Spectra of a He-Ne-Ar comparison lamp
were obtained after each exposure during the MMT observations, and 
spectra of a Hg-Ne-Ar comparison lamp were obtained after each exposure 
during the Keck II observations. They were used for
wavelength calibration. Two spectrophotometric standard stars, Feige 34 and 
HZ 44, were observed for flux calibration. 
The data reduction was made with the IRAF software package.
The two--dimensional spectra were bias--subtracted and 
flat--field corrected. Cosmic-ray removal, wavelength calibration, 
night sky background subtraction, correction for atmospheric extinction and 
absolute flux calibration were then performed.

One-dimensional spectra for the bright H {\sc ii} region {\it a} of \sbs\ 
were extracted from the Keck II and MMT data within large apertures of 
1\arcsec\ $\times$ 4\arcsec\ and 1\farcs5 $\times$ 3\arcsec\ respectively 
to minimize the effect of atmospheric dispersion (Fig. \ref{fig:brightsp}).  
In addition we extracted spectra showing hydrogen Balmer absorption lines
for four regions along  the main body of the galaxy (i.e. along the slit 
oriented at P.A. = --41$^{\circ}$).
The selected regions, denoted 1 to 4 (Keck II observations) and 1a to 4a
(MMT) in Tables \ref{t:emhahb} and \ref{t:abshdhg}
were at adjacent positions along 
the major axis of the BCD, with the origin taken to be at the center of
region {\it a}. Regions 1 and 1a coincide with region {\it c} in 
Fig. \ref{f1}a.
The spatial extents of the 4 regions along the slit were 3\farcs0, 2\farcs4, 
4\farcs8 and 3\farcs6 for the MMT observations and 3\farcs0, 2\farcs6, 
3\farcs6 and 4\farcs8 for the Keck II observations.
Additionally, spectra with apertures 10\farcs8 and 7\farcs2 
centered on region {\it c} were extracted from the two-dimensional 
MMT spectrum obtained at the position angle P.A. = 0$^\circ$ (regions 5 and 6 
in the Tables \ref{t:emhahb} and \ref{t:abshdhg}).

%


\begin{figure*}[hbtp]
 \hspace*{-0.0cm}\psfig{figure=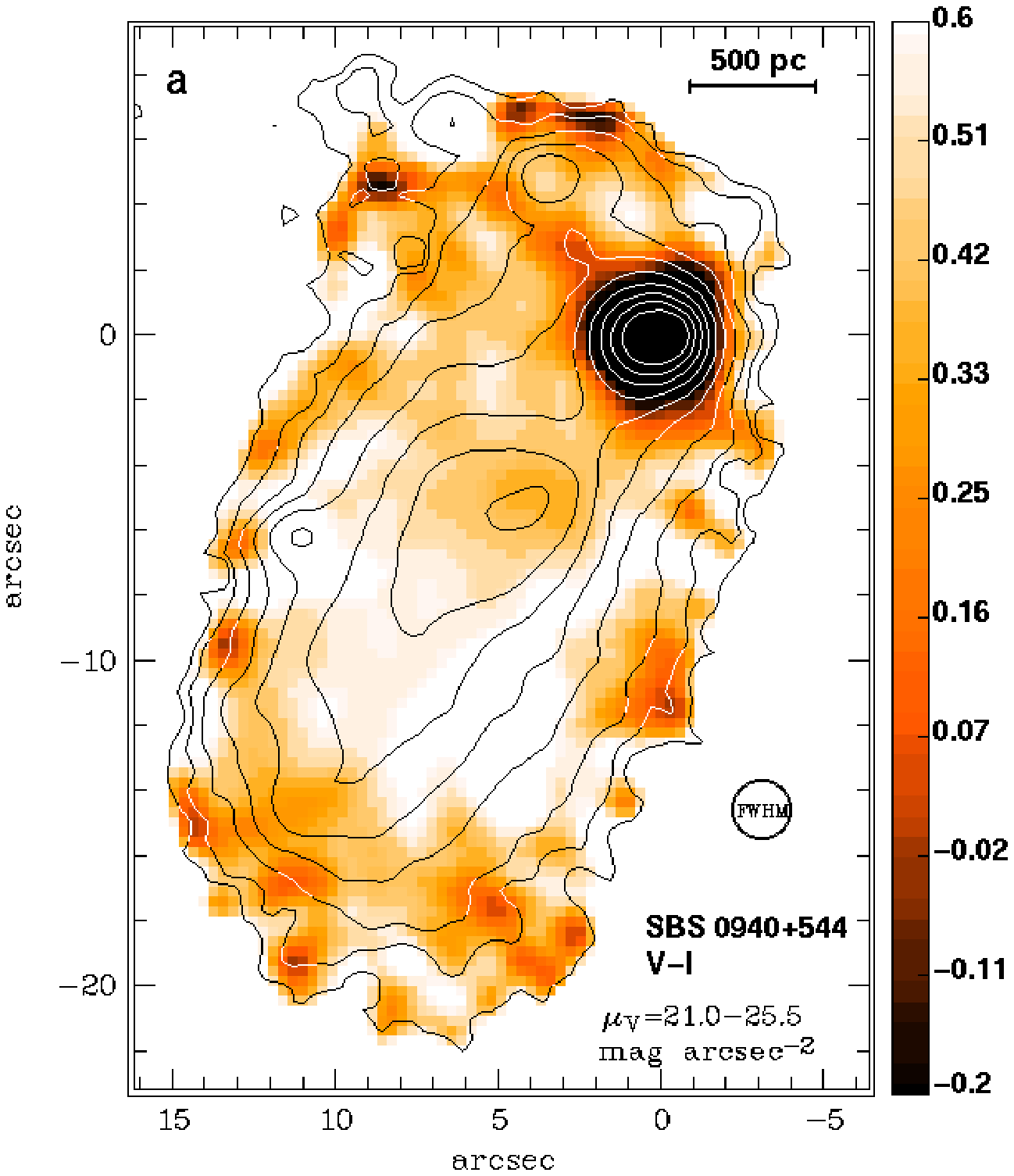,angle=0,width=9cm}
  \hspace*{-0.0cm}\psfig{figure=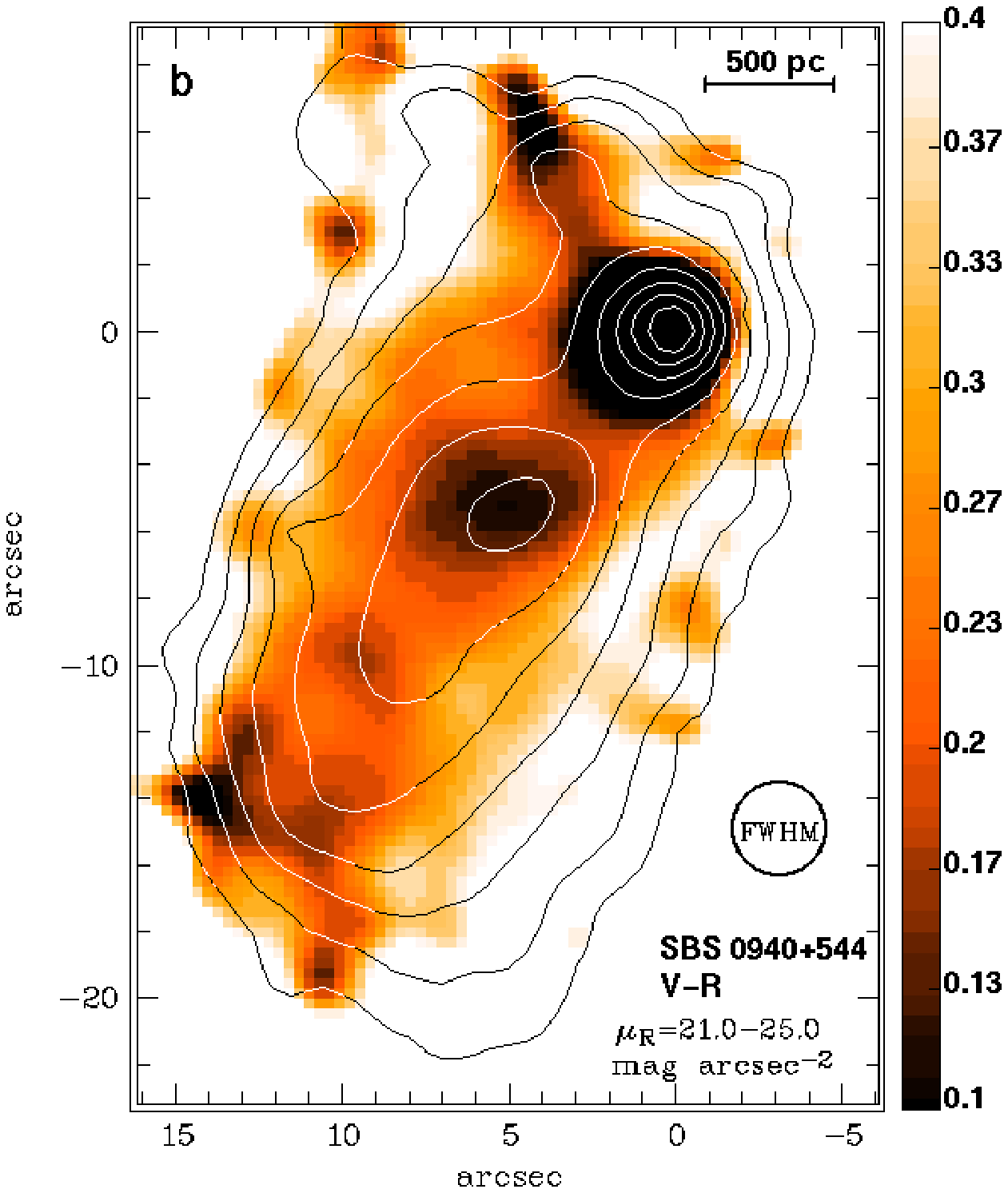,angle=0,width=9cm}
\caption[]{
(a) $V-I$ map of SBS 0940+544 displayed in the colour 
range --0.2 to +0.6 mag. The overlayed contours are from 21.0 to 25.5 
$V$ mag arcsec$^{-2}$  in steps of 0.5 mag. (b) $V-R$ map of the same 
region of the BCD displayed in the colour interval 0.1 to 0.4 mag.
The overlayed contours are from 21.0 to 25.0 $R$ mag arcsec$^{-2}$ 
in steps of 0.5 mag.
The bluest region in either map coincides with the northwestern 
H {\sc ii} region {\it a}, with an average $V-I$ and $V-R$ of 
$\la$ --0.5 mag and --0.2 mag, respectively. 
The angular resolution of each map is indicated by a
circle, the diameter of which is equal to the FWHM
of point sources in the corresponding smoothed input images
used for its derivation (1\farcs 8 and 2\farcs 75 for
the $V-I$ and $V-R$ map, respectively).}
\label{f2}
\end{figure*}

\section{Photometric analysis}

\subsection{Morphology and colour distribution \label{morph}}

Star formation activity in SBS 0940+544 takes place primarily 
in the unresolved high-surface-brightness H {\sc ii} region labeled {\it a}
in Fig. \ref{f1}a.
This  region with an absolute $V$ magnitude of --13.6 mag accounts for 
80\% of the total H$\alpha$ emission and for nearly 1/4 
of the total $V$ light of the BCD within the 25 $V$ mag arcsec$^{-2}$ isophote. 
A chain of faint point-like sources superposed on an 
asymmetric curved arm originating from region {\it a} is traceable to
$\sim$ 10\arcsec\ further to the northeast. This feature, labeled {\it b}
in Fig. \ref{f1}a, is not detected in H$\alpha$.  
Another feature is marginally visible to the SW side of the elongated 
low-surface-brightness (LSB) main body of the galaxy, at a surface brightness level 
$\ga$ 25.5 $V$ mag arcsec$^{-2}$ (region southwards of feature $d$ in Fig. \ref{f1}a).
Judging from the continuum-subtracted H$\alpha$ map (Fig. \ref{f1}b),
star formation in SBS 0940+544 is not exclusively confined to region 
{\it a} but occurs also in the LSB host. 
A secondary H$\alpha$-emitting region is visible close to the 
geometrical center of the outer isophotes (region {\it c} 
in Fig. \ref{f1}a). 

Evidence of recent star-formation activity along the major axis of 
the {\it cometary} LSB body of the BCD is also seen in the 
$V-I$ and $V-R$ colour maps (Fig. \ref{f2}). The $V-I$ and $V-R$ colours 
in region {\it c} are respectively $\sim$0.43 mag and $\sim$0.16 mag. 
The $V-R$ colour index remains nearly constant at
$\sim$0.2 mag over a large part of the LSB body
out to its southeastern extreme end
(position +13\arcsec,--13\arcsec\ in Fig. \ref{f2}b)
and increases to $\sim$0.34 mag at the
faint ($\mu_V\approx 24.0 - 24.5$ mag arcsec$^{-2}$)
southern region {\it d} (position
$\sim$ +5\arcsec, --14\arcsec).
The $V-I$ colour shows a similar general, although less well defined trend,
varying from $\sim$0.4-0.5 mag in the main body to $\sim$0.65 mag in 
region {\it d} and getting bluer again (0.36 mag) close to the 
southeastern tip of the LSB host.
The smooth colour change with decreasing intensity in the LSB body
is accompanied by a clear isophotal twist between 23.5 and 27 
$V$ mag arcsec$^{-2}$, with the position angle P.A. changing
from --40$^\circ$ to --10$^\circ$ and an average gradient of
8.5$^\circ$ mag$^{-1}$. 
Likewise, the ellipticity $10\times(1-\frac{b}{a})$ of the LSB host, 
$a$ and $b$ denoting respectively the semimajor and semiminor axis of 
the ellipse fitted to a given isophote, decreases 
from 5.5 at 24 $V$ mag arcsec$^{-2}$\ to 3.9 at 27 $V$ mag arcsec$^{-2}$.

Region {\it a} shows extraordinarily blue colours: $V-I$ = --0.5 mag and 
$V-R$ = --0.2 mag. Doublier et al. (\cite{Doublier97}) have obtained
$B$ and $R$ photometry of SBS 0940+544. They mistook 
region {\it a} for a Galactic foreground star.
Spectra of the object show without doubt that it is an extremely
compact H {\sc ii} region with very strong emission lines 
(Fig. \ref{fig:brightsp}) at the redshift of SBS 0940+544. Izotov et
al. (\cite{I91}) first derived element abundances in this H {\sc ii} region,
followed by Izotov et al. (\cite{ITL94}) who named it SBS 0940+544N.

\subsection{Surface photometry and colours}

\begin{figure*}[hbtp]
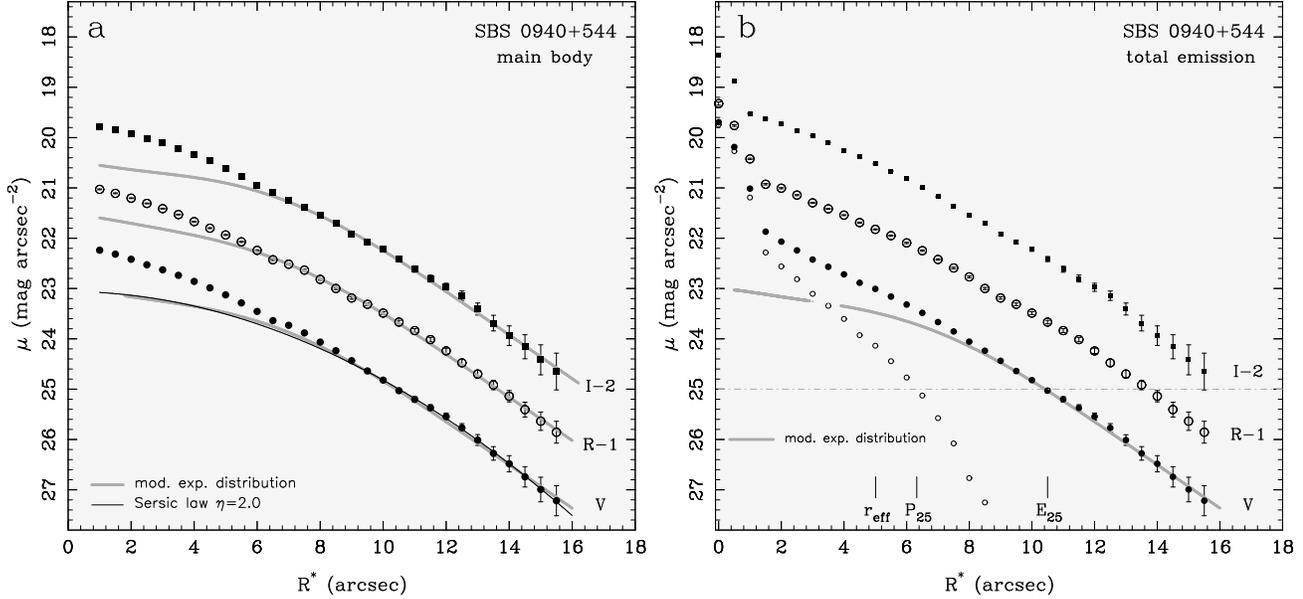

    \hspace*{-0.0cm}\psfig{figure=MS1281f3a.eps,angle=-90,width=8.5cm}
     \hspace*{-0.0cm}\psfig{figure=MS1281f3b.eps,angle=-90,width=8.5cm}
\caption[]{(a) Surface brightness profiles (SBPs) of the main body of 
SBS 0940+544
after subtraction of the brightest H {\sc ii} region {\it a} and the 
adjacent point sources {\it b} (cf. Fig. \ref{f1}a). 
For the sake of clarity, the $R$ and $I$ SBPs have been
shifted vertically by --1 and --2 mag respectively.
The modeled surface brightness distributions of the underlying LSB host 
in the $V$ band corresponding to a S\'ersic fitting law with an exponent 
$\eta=2$ (cf. Eq. \ref{eq:sersic}) and a modified exponential distribution with
($b$,$q$)=(3.0,0.9) (Eq. \ref{eq:p96a}) are shown respectively by the 
thin/black and thick/grey curves.  
(b) SBPs of the total emission of SBS 0940+544. The meaning of the symbols 
is the same as in the left panel. Small open circles show the surface
brightness distribution of the light in excess of the model for the LSB host
(thick/grey curve) in the $V$ band. The effective radius ${r_{\rm eff}}$ of 
the $V$ band SBP
and the isophotal radii $P_{25}$ and $E_{25}$ of the starburst and LSB 
component at 25 $V$ mag arcsec$^{-2}$\ are indicated.}
\label{f3}
\end{figure*}

As remarked in Sect. \ref{morph} (see also Sect. \ref{chem} and \ref{age}), 
signs of low-level ongoing or recent star formation 
are present all over the inner part of the main body of SBS 0940+544. 
In order to disentangle the stellar from the gaseous emission 
and better constrain the surface brightness and colour distribution of 
its faint underlying LSB host, we performed surface photometry in all 
available broad-band filters in two steps. 
First we fitted the intensity distribution and subtracted  both
the dominant region {\it a} and all 
adjacent point sources (designated by {\it b}, cf. Fig. \ref{f1}a).
We then derived surface brightness profiles (SBPs) for the main body 
using methods described in Papaderos et al. (\cite{Papa96b}). 
SBPs computed in this way (Fig.\ \ref{f3}a) trace the luminosity of the 
cometary LSB host of the BCD and the superimposed star-forming zone
extending from region {\it c} to the SE corner. 
The radial intensity distribution of the LSB host 
was approximated by adjusting different fitting laws to the outer part 
of the SBPs. We then computed SBPs for the whole BCD, and
by subtracting  from them the best model for 
the intensity distribution of the LSB host, we obtained the
surface brightness distribution of the discrete and diffuse starburst 
sources located in the northwestern part and main body (cf. Fig. \ref{f3}b).

%
%
%

The SBPs of the main body (Fig. \ref{f3}a) provide
an upper limit to the intensity distribution of the underlying unresolved older
stellar population of the LSB host, since part of the light 
for $R^*$ $\la$ 8\arcsec\ comes from region {\it c} and adjacent young sources. 
An exponential fitting law provides a reasonable approximation to the SBPs 
for $\mu_V\ga 24.5$ mag arcsec$^{-2}$, 
for radii $<$ 10\arcsec, however, it predicts
a higher intensity than the one observed.
This is also the case for the SBPs derived for the total
emission of the BCD (Fig. \ref{f3}b).
This type of convex profile with an exponential distribution
in the outer parts, and leveling 
off in the inner part (within 1--3 scale lengths) is not rare among
low-luminosity dwarf ellipticals (e.g. Binggeli \& Cameron \cite{binggeli91}), 
dwarf irregulars (Patterson \& Thuan \cite{patterson96}, Makarova \cite{m99}, 
van Zee \cite{v2000}) and blue low surface brightness galaxies 
(R\"onnback \& Bergvall 1994, Bergvall et al. 1999), and has been reported in a few BCDs 
(e.g. Vennik et al. \cite{vennik96},  Papaderos et al. \cite{Papa96b}, 
Telles et al. \cite{tmt97}, Papaderos et al. \cite{Papa99}, Fricke et
al. \cite{Fricke00}).
The observed profile shape suggests that a S\'ersic (\cite{S68}) profile
of the form
\begin{equation}
I(R^*) = I_0\,\exp\left(-\frac{R^*}{\alpha}\right)^{\eta}
\label{eq:sersic} 
\end{equation}
with $\eta$ $>$1 provides a better approximation to the light
of the LSB host than a pure exponential ($\eta$ = 1).

An alternative fitting formula which reproduces well an exponential 
distribution flattening inwards has been discussed by Papaderos et al. 
(\cite{Papa96b}): 
\begin{equation}
I(R^*) = I_0\,\exp\left( -\frac{R^*}{\alpha}\right)
\big[1-q\,\exp(-P_3(R^*))\big],
\label{eq:p96a} 
\end{equation}
where $P_3(R^*)$ is
\begin{equation}
P_3(R^*) = \left(\frac{R^*}{b\,\alpha}\right)^3+\left(\frac{R^*}{\alpha}\,\frac{1-q}{q}\right).
\label{eq:p96b} 
\end{equation}
Near the center, such an intensity distribution depends
on the relative central intensity depression $q=\Delta I/I_0$, where 
$I_0$ is the central intensity obtained by extrapolation
of the outer exponential slope with the scale length $\alpha$ to $R^*=0$, 
and $b \alpha$ is the cutoff--radius where the central flattening occurs.

\begin{figure}[hbtp]
    \hspace*{-0.0cm}\psfig{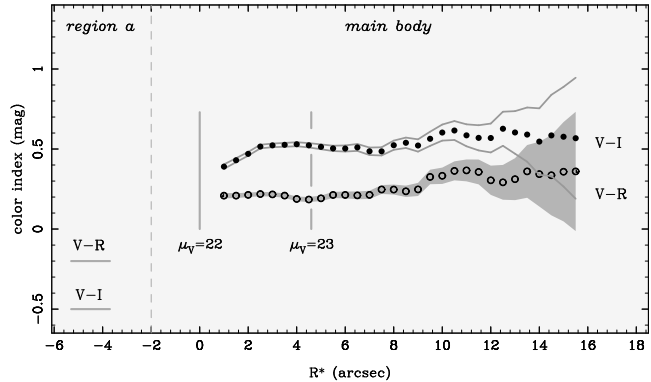}
\caption[]{$V-R$ and $V-I$ colour profiles of the main body of SBS 0940+544
derived by subtraction of the surface brightness profiles in
Fig. \ref{f3}a. The radial range corresponding to  the surface 
brightness interval between 22 and 23 $V$ mag arcsec$^{-2}$\ is indicated by 
vertical lines. The average colours of the bright
H {\sc ii} region {\it a} (cf. Fig. \ref{f1}a)  
( $V-R \sim $ --0.2 mag  and $V-I \la$ --0.5 mag ) 
are indicated by horizontal lines.}
\label{f4}
\end{figure}

\begin{figure}[hbtp]
    \psfig{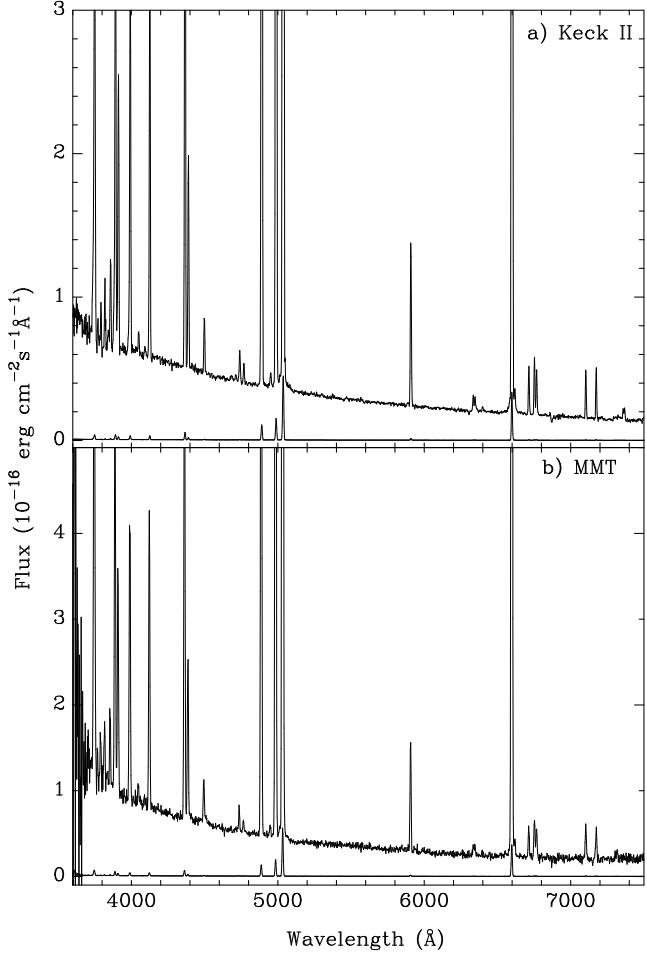}
    \caption{Spectra of the brightest H {\sc ii} region {\it a} in 
SBS 0940+544 obtained with: (a) the Keck II 
telescope and (b) the MMT. 
The lower spectra are the observed spectra downscaled by a 
factor of 100.}
    \label{fig:brightsp}
\end{figure}

Expressions (\ref{eq:sersic}) and (\ref{eq:p96a}) were fitted to the 
SBPs of the main body for radii $R^*\geq 9$\arcsec, where the colours become 
nearly constant (Fig. \ref{f4}). Fits with a S\'ersic profile give 
$\eta$ = 2.0 $\pm$ 0.1 and, as illustrated for the $V$ profile
in Fig. \ref{f3}a, are nearly indistinguishable from those 
using Eq. \ref{eq:p96a} with $b=3.0$ and $q=0.9$. The $V$ surface brightness 
distribution of the residual light in excess of the fit to the 
emission of the main body (Eq. \ref{eq:p96a})
is shown by small circles in the right panel of Fig. \ref{f3}.
 
Table \ref{photom} summarizes the derived photometric properties of the 
LSB host and the star-forming component of SBS 0940+544.
Cols.\ 2 \& 3 give respectively the central surface brightness 
$\mu_{E,0}$ and exponential scale length $\alpha\arcsec$ of the 
LSB host as obtained from linear fits to the SBPs for $R^*\geq 9$\arcsec\ and weighted by 
the photometric uncertainty of each point. 
The corresponding integrated magnitude $m_{\rm LSB}^{\rm fit}$ 
of a pure exponential distribution
$\mu_{E,0}-5\log(\alpha\arcsec)-2$ is listed in col. 4.
As discussed above, the surface brightness distribution of the LSB host becomes 
significantly shallower for $R^*<9$\arcsec\ than that predicted by inwards
extrapolation of the exponential slope observed at the outskirts of SBS 0940+544.
Therefore, for the case under study $m_{\rm LSB}^{\rm fit}$ overestimates the
apparent magnitude of the LSB host ($\approx m_{\rm LSB}$; col. 9).
Cols. 5 through 9 list quantities obtained from profile decomposition, 
with the intensity distribution of the LSB host modeled by 
Eq. \ref{eq:p96a}.
Cols.\ 5 and 7 give respectively the radial extent $P_{25}$ and $E_{25}$
of the starburst and LSB components, determined at a surface 
brightness level of 25\ mag arcsec$^{-2}$ (cf. Fig. \ref{f3}b).
The apparent magnitudes of each component within $P_{25}$ and $E_{25}$
are listed in cols.\ 6 and 8. Col.\ 9  gives the apparent magnitude of 
the LSB component in each band within the photometric radius 
of 15\farcs5, as derived from integration of the modeled 
distribution Eq. \ref{eq:p96a}. The total magnitude of the BCD
as inferred from SBP integration out to the same radius and by
summing up the flux within a polygonal aperture is given in
cols. 10 and 11, respectively. Col.\ 12 lists the effective 
radius $r_{\rm eff}$ and the radius $r_{80}$ which encircles 
80\% of the galaxy's total flux.
%
%
\begin{table*}
\caption{Structural properties of the starburst-- and LSB component 
of SBS 0940+544}
\label{photom}
\begin{tabular}{lccccccccccc}
\hline
Band & $\mu_{E,0}$ & $\alpha $ & $m_{\rm LSB}^{\rm fit}$ &
$P_{25}$  & $m_{P_{25}}$ & $E_{25}$ & $m_{
E_{25}}$ & $m_{\rm LSB}$  & $m_{\rm SBP}$ & $m_{\rm tot}$ & $r_{\rm eff}$,$r_{80}$  \\
          & mag arcsec$^{-2}$ &  arcsec   & mag    & kpc       &  mag             &  kpc
          &  mag             & mag      &   mag & mag & kpc \\
    (1) &   (2)         &   (3)     &  (4)      &   (5)            &
 (6)    &  (7)             &  (8)     &  (9)  & (10) & (11) & (12) \\
\hline
 $V$ & 20.46$\pm$0.12  & 2.51$\pm$0.05 & 16.46 & 0.82  & 17.89  & 1.36  & 17.53  & 17.34 &
 16.81$\pm$0.02  & 16.83 & 0.60,1.05\\
 $R$ & 20.14$\pm$0.10  & 2.52$\pm$0.05 & 16.13 & 0.76  & 18.34  & 1.46  & 17.1  & 16.97 &
 16.68$\pm$0.02  & 16.67 &  0.68,1.13\\
 $I$ & 19.83$\pm$0.13  & 2.50$\pm$0.06 & 15.85 & 0.83  & 17.89  & 1.54  & 16.87  & 16.76 &
 16.45$\pm$0.04 & 16.48 & 0.69,1.13\\
\hline
\end{tabular}
\end{table*}

The $V-R$ and $V-I$ colour profiles of the LSB host of SBS 0940+544 
as derived by subtraction of the corresponding SBPs (Fig. \ref{f3}a) 
are shown in Fig. \ref{f4}. Both colour profiles
show very weak gradients for $\mu_V\ga 23$ mag arcsec$^{-2}$, and become 
nearly constant for $R^*>9$\arcsec. 
Beyond that radius the colour profiles (Fig. 4) give 
average colour indices $V-R$= 0.33 $\pm$ 0.04 mag and
$V-I$ = 0.58 $\pm$ 0.03 mag for the LSB component. These values are
consistent with the integrated colours $V-R$ = 0.37 mag and
$V-I$ = 0.58 mag, determined from the apparent magnitudes of
the {\it modeled} intensity distribution of the LSB host within
$R^*=15$\arcsec\ (Table \ref{photom}, col. 9).

We have compared our results with those obtained for the $B$ and $R$ 
profiles by Doublier et al. (\cite{Doublier97}, \cite{Doublier99}). 
The agreement is not satisfactory.
The exponential scale length $\alpha$ of the LSB host derived here (Table \ref{photom})
is very similar in all bands ($V$, $R$ and $I$) being $\sim$2\farcs5 (320 pc).
Although Doublier et al. (\cite{Doublier97}) 
also report an exponential intensity decrease in $B$ and $R$ for radii 
$\ga 3$\arcsec, they derive for the distance assumed here
a $B$ and $R$ exponential scale length of $\sim$ 1 kpc, 
3 times greater than our value. 
A striking disagreement is also evident from a comparison of the 
colour profiles derived here and those presented in Doublier 
et al. (\cite{Doublier97}). 
The $B-R$ colour profile of Doublier et al. (\cite{Doublier97}) 
shows a maximum of $\sim$ 0.5 mag at $R^*\sim 0$\arcsec, then 
decreases monotonically with increasing radius to $\la$ 0.2 mag. 
Our colour profile shows the opposite trend: very blue colours at small radii
and nearly constant, moderately red $V-R$ and $V-I$ colours in the 
outskirts of SBS 0940+544. 

\begin{table*}[tbh]
\caption{Observed ($F$($\lambda$)), extinction-corrected ($I$($\lambda$))
fluxes and equivalent widths ($EW$) of the emission lines in the spectra 
of the brightest H {\sc ii} region in SBS 0940+544}
\label{t:Intens}
\begin{tabular}{lccrcccr} \hline \hline
  &\multicolumn{3}{c}{Keck II}&&\multicolumn{3}{c}{MMT} \\ \cline{2-4} 
\cline{6-8} \\
$\lambda_{0}$(\AA) Ion                  &$F$($\lambda$)/$F$(H$\beta$)
&$I$($\lambda$)/$I$(H$\beta$)&$EW$ (\AA)&&$F$($\lambda$)/$F$(H$\beta$)
&$I$($\lambda$)/$I$(H$\beta$)&$EW$ (\AA)   \\ \hline
3727\ [O {\sc ii}]             & 0.445 $\pm$0.007 &  0.471 $\pm$0.008&
62.0 $\pm$0.3&& 0.585 $\pm$0.008 &  0.587 $\pm$0.008&62.7 $\pm$0.7 \\
3750\ H12                      & 0.013 $\pm$0.001 &  0.022 $\pm$0.002&
 1.9 $\pm$0.2&& 0.022 $\pm$0.003 &  0.023 $\pm$0.004& 2.5 $\pm$0.3 \\
3771\ H11                      & 0.023 $\pm$0.001 &  0.032 $\pm$0.002&
 3.3 $\pm$0.2&& 0.044 $\pm$0.003 &  0.045 $\pm$0.005& 4.9 $\pm$0.4 \\
3798\ H10                      & 0.042 $\pm$0.001 &  0.052 $\pm$0.002&
 6.1 $\pm$0.2&& 0.059 $\pm$0.004 &  0.060 $\pm$0.005& 6.7 $\pm$0.5 \\
3820\ [He {\sc i}]             & 0.017 $\pm$0.001 &  0.018 $\pm$0.001&
 2.5 $\pm$0.3&&      ...         &        ...      &         ...~~~~  \\
3835\ H9                       & 0.058 $\pm$0.001 &  0.069 $\pm$0.002&
 8.6 $\pm$0.2&& 0.064 $\pm$0.004 &  0.065 $\pm$0.005& 7.4 $\pm$0.4 \\
3868\ [Ne {\sc iii}]           & 0.340 $\pm$0.005 &  0.356 $\pm$0.006&
49.6 $\pm$0.3&& 0.373 $\pm$0.006 &  0.374 $\pm$0.006&44.0 $\pm$0.7 \\
3889\ H8\ +\ He {\sc i}        & 0.184 $\pm$0.003 &  0.201 $\pm$0.004&
26.9 $\pm$0.3&& 0.204 $\pm$0.005 &  0.205 $\pm$0.006&25.1 $\pm$0.6 \\
3968\ [Ne {\sc iii}]\ +\ H7    & 0.259 $\pm$0.004 &  0.278 $\pm$0.005&
38.3 $\pm$0.3&& 0.288 $\pm$0.005 &  0.290 $\pm$0.006&39.8 $\pm$0.7 \\
4026\ He {\sc i}               & 0.014 $\pm$0.001 &  0.014 $\pm$0.001&
 2.1 $\pm$0.2&& 0.019 $\pm$0.004 &  0.019 $\pm$0.004& 2.7 $\pm$0.5 \\
4068\ [S {\sc ii}]             & 0.009 $\pm$0.001 &  0.009 $\pm$0.001&
 1.4 $\pm$0.2&& 0.007 $\pm$0.003 &  0.007 $\pm$0.003& 1.0 $\pm$0.4 \\
4101\ H$\delta$                & 0.244 $\pm$0.004 &  0.259 $\pm$0.004&
39.3 $\pm$0.3&& 0.267 $\pm$0.005 &  0.269 $\pm$0.006&40.0 $\pm$0.7 \\
4340\ H$\gamma$                & 0.461 $\pm$0.007 &  0.477 $\pm$0.007&
84.9 $\pm$0.4&& 0.487 $\pm$0.007 &  0.488 $\pm$0.007&86.9 $\pm$1.0 \\
4363\ [O {\sc iii}]            & 0.125 $\pm$0.002 &  0.128 $\pm$0.002&
23.2 $\pm$0.3&& 0.142 $\pm$0.004 &  0.142 $\pm$0.004&27.0 $\pm$0.7 \\
4388\ He {\sc i}               & 0.003 $\pm$0.001 &  0.003 $\pm$0.001&
 0.6 $\pm$0.2&&      ...         &        ...     &         ...~~~~    \\
4471\ He {\sc i}               & 0.036 $\pm$0.001 &  0.037 $\pm$0.001&
 7.4 $\pm$0.2&& 0.039 $\pm$0.003 &  0.039 $\pm$0.003& 7.3 $\pm$0.6 \\
4658\ [Fe {\sc iii}]           & 0.005 $\pm$0.001 &  0.005 $\pm$0.001&
 1.2 $\pm$0.3&& 0.003 $\pm$0.002 &  0.003 $\pm$0.002& 0.7 $\pm$0.5 \\
4686\ He {\sc ii}              & 0.006 $\pm$0.001 &  0.006 $\pm$0.001&
 1.3 $\pm$0.3&&      ...         &        ...     &         ...~~~~    \\
4711\ [Ar {\sc iv}]+He {\sc i} & 0.022 $\pm$0.001 &  0.022 $\pm$0.001&
 5.2 $\pm$0.3&& 0.020 $\pm$0.002 &  0.020 $\pm$0.002& 4.6 $\pm$0.5 \\
4740\ [Ar {\sc iv}]            & 0.013 $\pm$0.001 &  0.013 $\pm$0.001&
 3.0 $\pm$0.2&& 0.012 $\pm$0.003 &  0.012 $\pm$0.003& 2.9 $\pm$0.6 \\
4861\ H$\beta$                 & 1.000 $\pm$0.015 &  1.000 $\pm$0.015&
241.1 $\pm$0.8&& 1.000 $\pm$0.011 &  1.000 $\pm$0.011&250.1 $\pm$1.9 \\
4922\ He {\sc i}               & 0.010 $\pm$0.001 &  0.010 $\pm$0.001&
 2.4 $\pm$0.3&& 0.010 $\pm$0.002 &  0.010 $\pm$0.002& 2.6 $\pm$0.6 \\
4959\ [O {\sc iii}]            & 1.415 $\pm$0.021 &  1.402 $\pm$0.021&
348.8 $\pm$1.0&& 1.436 $\pm$0.014 &  1.435 $\pm$0.014&381.3 $\pm$2.3 \\
5007\ [O {\sc iii}]            & 4.251 $\pm$0.061 &  4.204 $\pm$0.061&
1052.0 $\pm$1.7&& 4.218 $\pm$0.034 &  4.215 $\pm$0.034&1154.2 $\pm$3.9 \\
5199\ [N {\sc i}]              & 0.003 $\pm$0.001 &  0.003 $\pm$0.001&
 0.9 $\pm$0.2&&      ...         &        ...     &         ...~~~~     \\
5271\ [Fe {\sc iii}]           & 0.004 $\pm$0.001 &  0.004 $\pm$0.001&
 1.2 $\pm$0.3&&      ...         &        ...     &         ...~~~~     \\
5876\ He {\sc i}               & 0.102 $\pm$0.002 &  0.097 $\pm$0.002&
39.0 $\pm$0.5&& 0.097 $\pm$0.003 &  0.097 $\pm$0.003&39.2 $\pm$1.1 \\
6300\ [O {\sc i}]              & 0.011 $\pm$0.001 &  0.010 $\pm$0.001&
 4.9 $\pm$0.4&& 0.007 $\pm$0.002 &  0.007 $\pm$0.002& 3.4 $\pm$0.7 \\
6312\ [S {\sc iii}]            & 0.012 $\pm$0.001 &  0.011 $\pm$0.001&
 5.3 $\pm$0.3&& 0.010 $\pm$0.002 &  0.010 $\pm$0.002& 4.7 $\pm$0.9 \\
6363\ [O {\sc i}]              & 0.004 $\pm$0.001 &  0.004 $\pm$0.001&
 1.9 $\pm$0.3&&      ...         &        ...     &         ...~~~~     \\
6563\ H$\alpha$                & 2.946 $\pm$0.043 &  2.750 $\pm$0.043&
1409.0 $\pm$2.5&& 2.761 $\pm$0.024 &  2.733 $\pm$0.025&1438.9 $\pm$6.0 \\
6584\ [N {\sc ii}]             & 0.013 $\pm$0.001 &  0.012 $\pm$0.001&
 3.9 $\pm$0.3&& 0.016 $\pm$0.002 &  0.016 $\pm$0.002& 8.6 $\pm$1.1 \\
6678\ He {\sc i}               & 0.029 $\pm$0.001 &  0.027 $\pm$0.001&
14.9 $\pm$0.5&& 0.027 $\pm$0.002 &  0.027 $\pm$0.002&14.3 $\pm$1.0 \\
6717\ [S {\sc ii}]             & 0.038 $\pm$0.001 &  0.035 $\pm$0.001&
19.2 $\pm$0.4&& 0.038 $\pm$0.002 &  0.038 $\pm$0.002&20.9 $\pm$1.3 \\
6731\ [S {\sc ii}]             & 0.030 $\pm$0.001 &  0.028 $\pm$0.001&
15.3 $\pm$0.4&& 0.026 $\pm$0.002 &  0.026 $\pm$0.002&14.7 $\pm$1.1 \\
7065\ He {\sc i}               & 0.026 $\pm$0.001 &  0.024 $\pm$0.001&
14.7 $\pm$0.4&& 0.032 $\pm$0.002 &  0.032 $\pm$0.002&19.8 $\pm$1.2 \\
7136\ [Ar {\sc iii}]           & 0.028 $\pm$0.001 &  0.026 $\pm$0.001&
16.5 $\pm$0.5&& 0.029 $\pm$0.002 &  0.029 $\pm$0.002&17.9 $\pm$1.2 \\
7320\ [O {\sc ii}]             & 0.007 $\pm$0.001 &  0.007 $\pm$0.001&
 4.6 $\pm$0.6&&      ...         &        ...     &         ...~~~~     \\
7330\ [O {\sc ii}]             & 0.007 $\pm$0.001 &  0.006 $\pm$0.001&
 4.1 $\pm$0.4&&      ...         &        ...     &         ...~~~~     \\
                     & & & & & \\
$C$(H$\beta$)\ dex             &\multicolumn {3}{c}{0.085 $\pm$0.019}&&\multicolumn {3}{c}{0.005 $\pm$0.011}  \\
$F$(H$\beta$)$^a$              &\multicolumn {3}{c}{1.21 $\pm$0.01}  &&\multicolumn {3}{c}{1.22 $\pm$0.01}    \\
$EW$(abs)~\AA                  &\multicolumn {3}{c}{1.1 $\pm$0.1}    &&\multicolumn {3}{c}{0.1 $\pm$0.3}      \\
\hline\hline
\end{tabular}

$^a$in units 10$^{-14}$\ erg\ s$^{-1}$cm$^{-2}$.
\end{table*}


\section{Chemical abundances \label{chem}}
 
Earlier determinations of element abundances in SBS 0940 +544 (Izotov et al.
\cite{I91}, \cite{ITL94}; Izotov \& Thuan \cite{IT98a}) have revealed a 
metallicity ranging from $Z_\odot$/31 to $Z_\odot$/25,
among the lowest for BCDs. Abundance determinations in this galaxy are 
therefore important for studying the origin of elements in low-metallicity 
environments and deriving the primordial helium abundance. We emphasize that
repeated observations of the same galaxy are very useful to estimate the 
robustness of the abundance determinations. This was one of the motivations for 
carrying out new MMT and Keck II observations of SBS 0940+544 and 
analyzing the abundance pattern in its brightest H {\sc ii} region. 
The spectra of region {\it a} used 
for the abundance determination are shown in Fig. \ref{fig:brightsp}.

A two-zone photoionized H {\sc ii} region has been assumed for 
abundance determination; the 
electron temperature $T_e$(O {\sc iii}) for the high-ionization region has 
been derived from the observed flux ratio 
[O {\sc iii}] $\lambda$\,4363/($\lambda$\,4959+5007), 
using a five-level atom model (Aller \cite{Aller84}) with atomic data from 
Mendoza (\cite{Mendoza83}). The electron temperature $T_e$(O {\sc ii})
for the low-ionization region has been derived using the empirical relation 
between $T_e$(O {\sc ii}) and $T_e$(O {\sc iii}) from the  
H {\sc ii} region photoionization models by Stasi\'nska (\cite{Stasinska90}).
The [S {\sc ii}]$\lambda$6717/$\lambda$6731 ratio was used to derive the 
electron number density $N_e$(S {\sc ii}). 

The spectra were corrected for interstellar extinction, where the extinction 
coefficient $C$(H$\beta$) was derived from the hydrogen Balmer 
decrement using the equations given in Izotov et al. 
(\cite{ITL94}) and the theoretical hydrogen emission line flux ratios from 
Brocklehurst (\cite{Brocklehurst71}).

The emission line fluxes were measured using a Gaussian profile 
fitting. The errors of the line fluxes measurements include the errors
in the fitting of profiles and those in the placement of the continuum. 
We also take into account the errors introduced by uncertainties in the spectral
energy distributions of standard stars. Standard star flux deviations for
both standard stars Feige 34 and HZ 44 are taken to be 1\%\ (Oke \cite{Oke90},
Bohlin \cite{Bohlin96}). These errors have been propagated in the
calculations of the element abundance errors.
The observed ($F$($\lambda$)) and extinction-corrected 
($I$($\lambda$)) emission line fluxes relative to the H$\beta$ emission line 
fluxes, the equivalent widths $EW$ of the emission lines,
the extinction coefficient $C$(H$\beta$), the observed flux
of the H$\beta$ emission line, and the equivalent width of the hydrogen 
absorption lines for region  {\it a}, derived from Keck II and MMT spectra
with the slit position P.A. = --41$^\circ$ are listed in Table \ref{t:Intens}.
The derived extinction coefficient $C$(H$\beta$) is small.

\begin{table}[tbh]
\caption{Physical conditions and element abundances in the brightest 
H {\sc ii} region of SBS 0940+544}
\label{t:Chem}
\begin{tabular}{lcc} \hline \hline
Value                               & Keck II     & MMT \\ \hline
$T_e$(O {\sc iii})(K)                     &18830$\pm$220&20000$\pm$370 \\
$T_e$(O {\sc ii})(K)                      &15320$\pm$170&15690$\pm$260 \\
$T_e$(S {\sc iii})(K)                     &17330$\pm$190&18300$\pm$300 \\
$N_e$(S {\sc ii})(cm$^{-3}$)              &   170$\pm$60&    10$\pm$10\\
$N_e$(He {\sc ii})(cm$^{-3}$)             &    10$\pm$10&    60$\pm$10 \\
$\tau$($\lambda$3889)               &       0.0   &    0.5$\pm$0.3 \\ \\
O$^+$/H$^+$($\times$10$^5$)         &0.389$\pm$0.013&0.447$\pm$0.020\\
O$^{+2}$/H$^+$($\times$10$^5$)      &2.727$\pm$0.081&2.438$\pm$0.102\\
O$^{+3}$/H$^+$($\times$10$^6$)      &0.201$\pm$0.033&      ...      \\
O/H($\times$10$^5$)                 &3.137$\pm$0.082&2.885$\pm$0.116\\
12 + log(O/H)                       &7.497$\pm$0.011&7.460$\pm$0.018\\ \\
N$^{+}$/H$^+$($\times$10$^7$)       &0.874$\pm$0.034&1.115$\pm$0.112\\
ICF(N)$^a$                          &8.05\,~~~~~~~~~~&6.46\,~~~~~~~~~~\\
log(N/O)                            &--1.649$\pm$0.021~~&--1.603$\pm$0.048~~\\ \\
Ne$^{+2}$/H$^+$($\times$10$^5$)     &0.476$\pm$0.015&0.436$\pm$0.020\\
ICF(Ne)$^a$                         &1.15\,~~~~~~~~~~&1.18\,~~~~~~~~~~\\
log(Ne/O)                           &--0.758$\pm$0.018~~&--0.748$\pm$0.028~~\\ \\
S$^+$/H$^+$($\times$10$^7$)         &0.608$\pm$0.015&0.583$\pm$0.031\\
S$^{+2}$/H$^+$($\times$10$^7$)      &3.643$\pm$0.213&2.794$\pm$0.516\\
ICF(S)$^a$                          &2.10\,~~~~~~~~~~&1.83\,~~~~~~~~~~\\
log(S/O)                            &--1.546$\pm$0.025~~&--1.670$\pm$0.069~~\\ \\
Ar$^{+2}$/H$^+$($\times$10$^7$)     &0.744$\pm$0.021&0.751$\pm$0.050\\
Ar$^{+3}$/H$^+$($\times$10$^7$)     &1.124$\pm$0.069&0.959$\pm$0.205\\
ICF(Ar)$^a$                         &1.01\,~~~~~~~~~~&1.02\,~~~~~~~~~~\\
log(Ar/O)                           &--2.219$\pm$0.020~~&--2.219$\pm$0.059~~\\ \\
Fe$^{+2}$/H$^+$($\times$10$^7$)     &0.934$\pm$0.212&0.547$\pm$0.384\\
ICF(Fe)$^a$                         &10.1\,~~~~~~~~~~&8.07\,~~~~~~~~~~\\
log(Fe/O)                           &--1.537$\pm$0.099~~&--1.815$\pm$0.313~~\\
$[$O/Fe$]$                          &0.103$\pm$0.099&0.395$\pm$0.313\\ \\
He$^+$/H$^+$($\lambda$4471)         &0.0801$\pm$0.0023&0.0829$\pm$0.0066\\
He$^+$/H$^+$($\lambda$5876)         &0.0821$\pm$0.0015&0.0805$\pm$0.0024\\
He$^+$/H$^+$($\lambda$6678)         &0.0813$\pm$0.0023&0.0804$\pm$0.0055\\
He$^+$/H$^+$(mean)                  &0.0815$\pm$0.0011&0.0807$\pm$0.0021\\
He$^{+2}$/H$^+$($\lambda$4686)      &0.0005$\pm$0.0001&       ...       \\
He/H                                &0.0820$\pm$0.0011&0.0807$\pm$0.0021\\
$Y$                                 &0.2468$\pm$0.0034&0.2439$\pm$0.0066\\ \hline
\end{tabular}

$^a$ICF is the ionization correction factor.
\end{table}


The ionic abundances of O$^{+2}$, Ne$^{+2}$, Ar$^{+3}$, He$^+$ and 
He$^{+2}$ were obtained using the electron temperature $T_e$(\ion{O}{iii}).
The electron temperature $T_e$(\ion{O}{ii}) is adopted to calculate the O$^+$, 
N$^+$, S$^+$ and Fe$^{+2}$ ionic abundances, while the intermediate value of 
the electron temperature $T_e$(\ion{S}{iii}) was used to derive the ionic 
abundances of Ar$^{+2}$ and S$^{+2}$ (Garnett \cite{Garnett92}).

The total heavy element abundances were obtained using ionization correction 
factors (ICF) following Izotov et al. (\cite{ITL94}, \cite{Izotov97}) and 
Thuan et al. (\cite{til95}). 
For oxygen the total abundance is the sum of O$^+$,
O$^{+2}$ and O$^{+3}$ ion abundances. Although emission lines of O$^{+3}$
are absent in the optical spectrum, this ion resides in the He$^{+2}$ region.
Therefore, we derive the O$^{+3}$ ion abundance using 
the He {\sc ii} $\lambda$4686 \AA\ emission line intensity as described by
Izotov \& Thuan (\cite{IT99}).  
The ionic and heavy element abundances for the  brightest \ion{H}{ii} region 
in SBS 0940+544 together with electron temperatures 
and electron number densities 
are given in Table \ref{t:Chem} along with 
the adopted ionization correction factors.

The oxygen abundance derived from the Keck II
and MMT spectra are 12 + log(O/H) = 7.50 $\pm$ 0.01 and 7.46 $\pm$ 0.02 
respectively. For comparison, Izotov et al. (\cite{I91}) derived 
12 + log(O/H) = 7.52 $\pm$ 0.12, while Izotov et al. (\cite{ITL94}) obtained
12 + log(O/H) = 7.37 $\pm$ 0.02. The former value was calculated using a 
three-level atom model. If the five-level atom model is adopted instead then 
the oxygen abundance is decreased by $\sim$ 0.04 dex (Izotov \& Thuan 
\cite{IT99}). The latter value is significantly lower than the previous ones.
However, Izotov \& Thuan (\cite{IT98a})
noted that Izotov et al. (\cite{ITL94}) used an erroneously low [O {\sc iii}]
$\lambda$5007 emission line intensity. They rederived the oxygen abundance 
using the data of Izotov et al. (\cite{ITL94}) and found
12 + log(O/H) = 7.43 $\pm$ 0.01. Thus, all 12 + log (O/H) values are confined 
in the narrow range 7.43 -- 7.50, or $Z_\odot$/31 -- 
$Z_\odot$/26.\footnote{12 + log(O/H)$_{\odot}$ = 8.92 (Anders \& Grevesse 
\cite{Anders89}).}

The ratios of other heavy element abundances to the oxygen abundance derived 
from the Keck II and MMT observations (Table \ref{t:Chem}) are
mutually consistent and they are also in agreement with 
the heavy element abundance
ratios derived earlier in SBS 0940+544 and in other low-metallicity
BCDs (Izotov \& Thuan \cite{IT99}).

The very low metallicity of SBS 0940+544 and the presence of very strong 
emission lines in its spectrum makes it one of the best galaxies for the
determination of the primordial helium abundance. For this, we use 
the five strongest He {\sc i} $\lambda$3889, $\lambda$4471, $\lambda$5876, 
$\lambda$6678 and $\lambda$7065 emission lines in both Keck II and MMT spectra. 
The first and last
He {\sc i} emission lines are more sensitive to collisional and fluorescent 
enhancement mechanisms. They are used to correct other He {\sc i} 
emission lines for these effects (Izotov et al. \cite{ITL94}, \cite{Izotov97}). 
The helium abundance is derived from the corrected intensities of the 
He {\sc i} $\lambda$4471, $\lambda$5876, $\lambda$6678 emission lines
and is shown in Table \ref{t:Chem}. The mean values of the $^4$He mass
fraction $Y$ = 0.247 $\pm$ 0.003 (Keck II) and $Y$ = 0.244 $\pm$ 0.006 (MMT) 
(see Table \ref{t:Chem}) are consistent with the previously derived value $Y$ = 
0.247 $\pm$ 0.007 (Izotov \& Thuan \cite{IT98a})
and they are close to the primordial $^4$He mass fraction
$Y_p$ = 0.244 $\pm$ 0.002 derived by extrapolating the $Y$ vs O/H linear 
regression to O/H = 0 (Izotov \& Thuan \cite{IT98a}), or to 
$Y_p$ = 0.245 $\pm$ 0.002 derived in the two most metal-deficient BCDs I Zw 18 
and SBS 0335--052 (Izotov et al. \cite{ICFGGT99}). 

   It was pointed out in Sect. \ref{obsspec} that atmospheric dispersion may
introduce some uncertainties in the abundance determination. To estimate these
uncertainties we measure the fluxes of the [O {\sc ii}] 
$\lambda$3727 and [O {\sc iii}] $\lambda$4363 emission lines in an aperture
1\arcsec$\times$1\arcsec\ placed on the maximum in the flux distribution,
and those in an aperture with the same size but displaced along the slit
by 0\farcs4 and 0\farcs1 respectively for the [O {\sc ii}] and [O {\sc iii}]
lines. We obtain relative flux differences of $\sim$ 7\% and $\sim$ 1\% for
the [O {\sc ii}] and [O {\sc iii}] emission lines. 
The effect of the atmospheric dispersion results in an
uncertainty not exceeding 1\% of the total oxygen abundance, because oxygen
in the H {\sc ii} region is mainly O$^{+2}$ (see Table \ref{t:Chem}).
However, because the H {\sc ii} region is much more extended than the size
of the aperture used for analysis of the atmospheric dispersion, 
and of the large apertures used in the extraction of spectra for the abundance 
analysis, the effect of atmospheric dispersion is expected
to be smaller for the [O {\sc ii}] $\lambda$3727 emission line and negligible
compared to the statistical errors for the [O {\sc iii}] $\lambda$4363 
emission line. 
Therefore, we decided not to include it in the error budget.


\begin{figure}[hbtp]
    \psfig{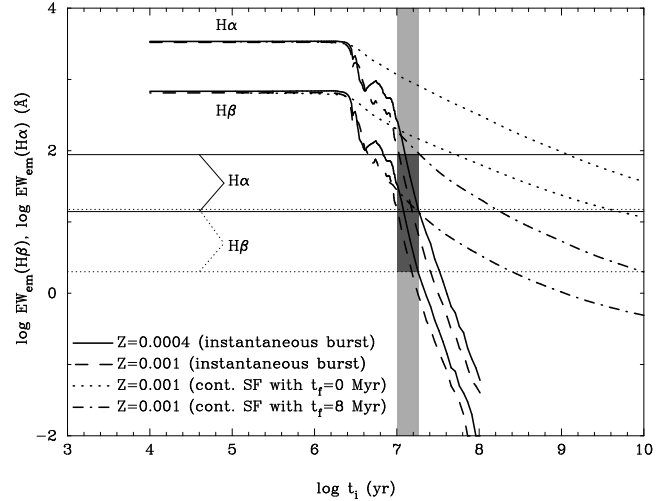}
    \caption{Equivalent widths of the nebular emission
     lines H$\alpha$ and H$\beta$ vs. age for 
     an instantaneous burst and a heavy element mass fraction 
     $Z$ = 1/20 $Z_\odot$ (thick dashed lines) 
     and $Z$ = 1/50 $Z_\odot$ (thick solid lines) calculated with the 
PEGASE.2 code (Fioc \& Rocca-Volmerange \cite{F97}). 
Model predictions are also shown for the case of constant continuous 
star formation starting at an age defined by 
the abscissa $t_{\rm i}$ and stopping at $t_{\rm f}$, with 
$t_{\rm f}$ = 0 Myr (dotted lines), and $t_{\rm f}$ = 8 Myr (dash-dotted 
lines).
     Thin solid and dotted horizontal lines show the upper and lower
     limits of observed $EW$(H$\alpha$)
     and $EW$(H$\beta$) for 4 regions of the underlying galaxy.
     The shaded rectangular region shows the age limits in the case
     of an instantaneous burst. 
     }
    \label{fig:ewhbha}
\end{figure}



\section{Age of the underlying stellar population \label{age}}

One of the key problems debated in recent years is the evolutionary status of
very metal-deficient BCDs: are these galaxies young or old? To resolve this 
problem an analysis of the stellar populations is required. 
The metallicity of the galaxy is one of the key parameters that determines the
photometric properties of stellar populations.
Izotov \& Thuan (\cite{IT99}), considering the abundance ratios of the heavy
elements in BCDs, have suggested that the BCDs with 12 + log (O/H)
$\la$ 7.6 might be young unevolved galaxies. SBS 0940+544 fulfills this 
abundance condition.
Furthermore,  the nitrogen-to-oxygen abundance ratios log(N/O) = 
--1.65 $\pm$ 0.02 and --1.60 $\pm$ 0.05 derived respectively from the Keck II 
and MMT spectra (Table~\ref{t:Chem}) are confined to the 
narrow range obtained earlier by Thuan et al. (\cite{til95}) and Izotov \& 
Thuan (\cite{IT99}) for extremely metal-poor BCDs.
The constancy of the N/O abundance ratio in these BCDs
with a very small spread of values around the mean suggests that the production
of primary nitrogen occurs in massive stars only and hence these systems 
are likely
to be young, since intermediate-mass stars have not had time to release their
nucleosynthetic products. 
On the other hand, some recent chemical evolution models have been 
succesful in reproducing the observed constancy of the N/O ratio, 
with nitrogen produced by the longer-lived intermediate-mass stars 
(Pilyugin \cite{Pil99}; Henry et al. \cite{H00}).
 
Deep $V, R, I$, and H$\alpha$ imaging, high signal-to-noise ratio spectra 
showing strong emission lines and the detection of hydrogen Balmer absorption
lines in the main body of SBS 0940+544 enable the study of stellar populations
and constraint of their ages through various techniques.

The light of the brightest H {\sc ii} region 
of the galaxy is dominated by a very young stellar population.
The very blue continuum and large $EW$(H$\beta$) = 241 -- 250 \AA\ 
(Fig. \ref{fig:brightsp}, Table \ref{t:Intens}) are consistent with an age 
of an instantaneous burst not 
exceeding 4 Myr, for a metallicity $Z$ = 1/20~-- 1/50 $Z_\odot$ 
(Fig.~\ref{fig:ewhbha}). However, despite that young age, no Wolf-Rayet
stars are detected in our high signal-to-noise ratio spectra. We also note
that the nebular He {\sc ii} $\lambda$4686 emission line is not detected
in the MMT spectrum and its intensity in the Keck II spectrum
is only 0.6\% of H$\beta$, or several
times weaker than in the very metal-deficient BCDs I Zw 18, SBS 0335--052 and 
Tol 1214--277 with detected Wolf-Rayet stellar populations (Izotov et al.
\cite{I97}; Izotov et al. \cite{ICFGGT99}; Fricke et al. \cite{Fricke00}).

      The interpretation of the photometric data for the brightest 
H {\sc ii} region is not straightforward because of the strong 
contamination of stellar radiation by the emission of the ionized gas.
The colour $V-I$ $\sim$ --0.5 mag is too blue to be explained even by the 
youngest stellar population, because $\sim$ 40\% of the $V$ light is
contributed by the very strong oxygen emission lines [O {\sc iii}] 
$\lambda$4959, $\lambda$5007. Therefore broad-band photometry
alone is not sufficient to constrain the age of young
star-forming regions. However, spectroscopic data in conjunction with 
photometric data allows us to disentangle stellar from gaseous emission.

Data for bright H {\sc ii} regions do not permit the detection of the small 
fraction of light contributed by an old stellar population 
(e.g., Papaderos et al. \cite{Papa98}; Fricke et al. \cite{Fricke00}; 
Noeske et al. \cite{Noeske00}). 
The old population in such regions, if present,
is hidden by the strong emission from the young stars and the gas ionized by 
the numerous massive stars in the star-forming region. More promising 
for the detection of an old stellar population is the study of the 
underlying extended emission from the host galaxy as gaseous contamination
is minimized there. Nearly all BCDs show an underlying stellar component 
with red colours, consistent with ages greater than a few Gyr 
(Loose \& Thuan \cite{Loose86};  Papaderos et al. 
\cite{Papa96a,Papa96b}; Telles \& Terlevich \cite{telles97}). However, the 
colours of the extended emission in the few very low metallicity BCDs 
are consistent with younger ages.

Another method for deriving ages is by fitting the observed galaxy spectral 
energy distribution (SED) with theoretical SEDs for various stellar
population ages and star formation histories. This method has been applied 
to several extremely metal-deficient BCDs with $Z$ = (1/20 -- 1/40)$Z_\odot$ 
(e.g., SBS 0335--052 (Papaderos et al. \cite{Papa98}), 
SBS 1415+437 (Thuan et al. \cite{TIF99}) 
and Tol 1214--277 (Fricke et al. \cite{Fricke00})). 
The SED of the underlying stellar component in these galaxies, obtained after 
subtraction of the ionized gas emission from the observed spectrum,
could be fitted by a stellar population not older than a few hundred Myr.

However, both methods are subject to uncertainties introduced by 
extinction. Therefore, other less extinction-dependent methods
are desirable to better constrain the age of stellar populations. 
We use in this Section two such methods. One relies on the Balmer nebular 
emission line equivalent widths and the other on the Balmer stellar 
absorption line equivalent widths. Such an analysis is feasible because of the 
exceptionally high-quality of the Keck II and MMT spectra. 
Because age estimates from all above mentioned methods 
depend on the adopted star formation history in the galaxy, we
consider next different scenarios of star formation (instantaneous burst,
continuous star formation) with varying extinction and star formation 
rate (SFR).
We then put together the constraints imposed on the stellar age and 
extinction by all four methods to draw a consistent picture for 
the populations in SBS 0940+544.

\subsection{The case of an instantaneous burst \label{inst}}

\subsubsection{Age from the nebular emission lines
\label{sec:emission}} 

 Under the assumption of an ionization-bounded H {\sc ii} region, the strongest 
hydrogen recombination emission lines H$\alpha$ and H$\beta$ provide an 
estimate for the age of the young stellar population when 
late O and early B stars are still present.
The ionizing flux from such a young stellar population
and hence the equivalent widths of the Balmer emission lines have a very 
strong temporal evolution. 

Fluxes and equivalent widths of the H$\alpha$ and 
H$\beta$ emission lines are obtained for all four regions in the main body.
Because the H$\beta$ emission line is narrower than the absorption line in
these regions and does not fill the absorption component, its flux was measured
using the continuum level at the bottom of the absorption line.
This level has been chosen by visually interpolating from the absorption line
wings to the center of the line.
The extinction coefficient $C$(H$\beta$)
in those regions is derived from the  H$\alpha$/H$\beta$ flux ratio.
For this, the theoretical recombination H$\alpha$/H$\beta$ flux ratio of
2.8 is adopted, which is typical for hot low-metallicity H {\sc ii} regions.
No correction for the absorption line equivalent width has been made in this
case.
Results of the measurements together with errors are shown in Table 
\ref{t:emhahb}. In Fig.~\ref{fig:ewhbha} we compare the 
measured H$\alpha$ and H$\beta$ emission line equivalent widths with those
predicted at a given age of an instantaneous burst.
The time 
evolution of the H$\alpha$ and H$\beta$ emission line equivalent widths for 
the heavy element mass fractions $Z_\odot$/50 ( thick solid
lines) and $Z_\odot$/20 (thick dashed lines) is calculated using the galactic evolution 
code PEGASE.2
(Fioc \& Rocca-Volmerange \cite{F97}). This code is based on the Padua 
isochrones (Bertelli et al. \cite{Bertelli94}) and stellar atmosphere
models from Lejeune et al. (\cite{Lejeune98}). 
Thin solid and dotted horizontal lines indicate respectively the highest and 
lowest observed equivalent widths of H$\alpha$ and H$\beta$ for the 4 regions 
of the underlying galaxy. 
The shaded region shows that the ages range between 10 and 20 Myr, meaning 
that the gas in the main body is likely ionized by early B stars rather than 
by O stars.

   This age estimate is valid if the population of
ionizing stars is large enough and can be described by an initial
mass function (IMF) with an analytical form, for example a Salpeter IMF. 
However, in the case of the main
body of SBS 0940+544, the number of ionizing stars is small and stochastic 
effects in the distribution of stars of different masses might be important. 
In particular, because of stochastic star formation, the ionization
may be caused by a few O stars instead of a population of B stars. Taking the 
number of ionizing Lyc photons to be respectively 10$^{49}$ s$^{-1}$ 
and 10$^{48}$ s$^{-1}$ for a O7{\sc v} and a B0{\sc v} star (Vacca, Garmany 
\& Shull \cite{Vacca96}), we find from the observed fluxes of the H$\beta$ 
emission line that the number of O7{\sc v} and B0{\sc v} stars is in 
the range 2 -- 7 and 23 -- 70, respectively, depending on the location
in the main body (see Table \ref{t:emhahb}). Assuming 
a Salpeter IMF with a slope 2.35, and upper and lower star mass limits of 120 
$M_\odot$ and 2 $M_\odot$, we derive a total mass 10$^4$ -- 10$^5$ $M_\odot$ 
for the young stellar population. 
Cervi\~no, Luridiana \& Castander 
(\cite{Cervino00}) have analyzed the stochastic nature of the IMF in young
stellar clusters with solar metallicity. In particular, they find that
in the range of equivalent widths $EW$(H$\beta$) = 1 -- 10 \AA, the dispersion
of age at fixed $EW$(H$\beta$) is not greater than 5\% - 10\% if the total
mass of the cluster lies in the range 10$^4$ -- 10$^5$ $M_\odot$.
Hence, we argue that stochastic effects in SBS 0940+544
do not significantly change our age estimate from the hydrogen 
emission line equivalent widths.


\begin{figure}[hbtp]
    \psfig{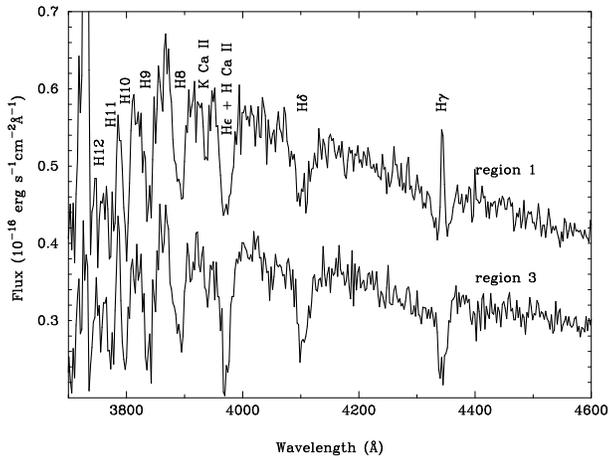}
\caption{Keck II spectra of regions 1 and 3 
(respectively at 6\farcs3 and 12\farcs2 from the brightest H {\sc ii} region)
showing H$\gamma$, H$\delta$ and other hydrogen and Ca {\sc ii} absorption
lines.
     }
    \label{fig:absorb}
\end{figure}



\begin{figure}[hbtp]
    \psfig{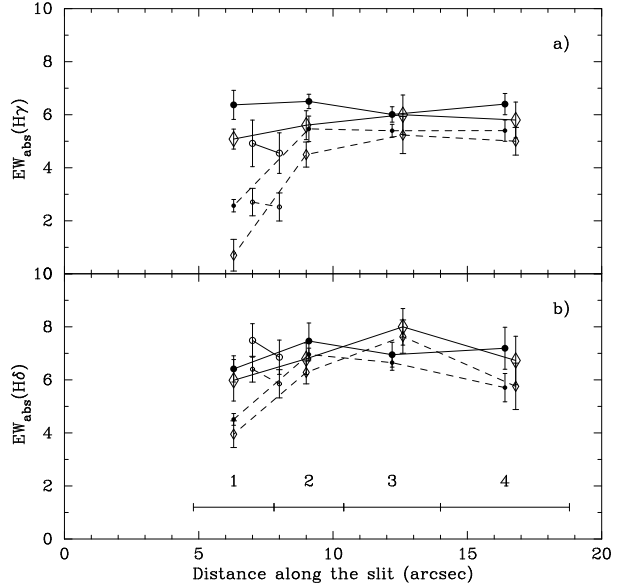}
    \caption{The equivalent widths of H$\gamma$ (a) and H$\delta$ (b) 
     absorption lines in 4 regions along the main 
     body of SBS 0940+544. Filled large circles, open large rhombs and solid 
     lines show the corrected $EW_{\rm abs}$(H$\gamma$) and 
     $EW_{\rm abs}$(H$\delta$) obtained respectively
     from the Keck II and MMT observations with position angle
     of the slit P.A. = --41$^{\circ}$. Large open circles
     connected by solid lines show the data from the MMT spectrum with
     the slit oriented at P.A. = 0$^{\circ}$.
     Small symbols and dashed lines denote the equivalent widths 
     uncorrected for emission.
     } 
    \label{fig:hdhg}
\end{figure}


\begin{figure}[hbtp]
    \psfig{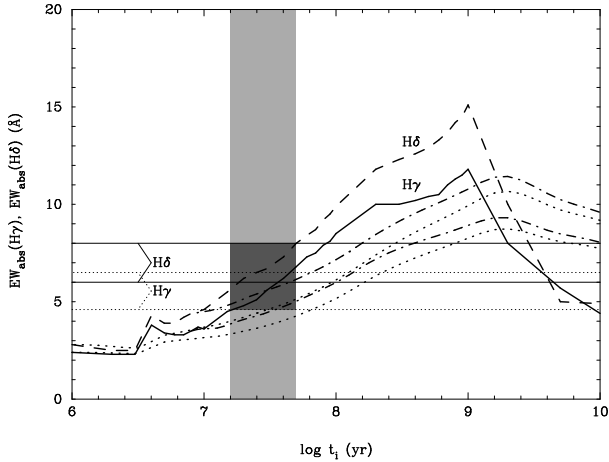}
    \caption{Equivalent widths of 
    H$\gamma$ and H$\delta$ absorption lines vs age for an instantaneous 
    burst with metallicity $Z$ = 1/20 $Z_\odot$ from Gonzalez Delg\'ado et al.
    (\cite{GonLeith99b}) (for ages $\leq$ 1 Gyr) and Bica \& Alloin 
    (\cite{Bica86}) (for ages $>$ 1 Gyr), shown by thick solid and dashed 
    lines respectively. The upper and lower limits of measured $EW$(H$\gamma$)
    and $EW$(H$\delta$) in 4 regions along 
    the main body of the galaxy are shown by 
    thin solid and dotted horizontal lines.
Model predictions are also shown for the case of constant continuous 
star formation starting at an age defined by 
the abscissa $t_{\rm i}$ and stopping at $t_{\rm f}$, with 
$t_{\rm f}$ = 0 Myr (dotted lines), and $t_{\rm f}$ = 8 Myr (dash-dotted 
lines).
    The shaded rectangular region gives the age limits in the case of an
instantaneous burst.
     }
    \label{fig:abshghd}
\end{figure}





\begin{table*}[tbh]
\caption{Fluxes, equivalent widths of the H$\alpha$ and H$\beta$ emission 
lines and the extinction coefficient $C$(H$\beta$) in different regions 
of the SBS 0940+544 main body}
\label{t:emhahb}
\begin{tabular}{lcrcrrrrrc} \hline \hline
Telescope&\# of region&Distance$^a$&Aperture$^b$&& 
$F$(H$\alpha$)$^c$  &$F$(H$\beta$)$^c$  &$EW$(H$\alpha$)$^d$  
&$EW$(H$\beta$)$^d$ &$C$(H$\beta$)   \\ \hline

Keck II$^e$ &1& 6.3~~~~~~&1.0$\times$3.0 && 13.30 $\pm$0.14 & 3.91 $\pm$0.12 & 67.18 $\pm$0.75 & 12.37 $\pm$0.24 & 0.13 $\pm$0.03 \\
     &2& 9.1~~~~~~&1.0$\times$2.6 && 3.15 $\pm$0.08 & 0.72 $\pm$0.08 & 22.23 $\pm$0.56 & 3.26  $\pm$0.24 & 0.29 $\pm$0.05 \\
     &3& 12.2~~~~~~&1.0$\times$3.6 && 2.06 $\pm$0.10 & 0.43 $\pm$0.08 & 13.84 $\pm$0.64 & 2.18  $\pm$0.31 & 0.35 $\pm$0.07 \\
     &4& 16.4~~~~~~&1.0$\times$4.8 && 3.60 $\pm$0.13 & 1.05 $\pm$0.12 & 45.72 $\pm$1.71 & 9.15  $\pm$0.73 & 0.13 $\pm$0.05 \\
 
  \hline

MMT$^e$ &1a& 6.3~~~~~~&1.5$\times$3.0 && 14.91 $\pm$0.39 & 4.22 $\pm$0.29 & 88.14 $\pm$2.32& 14.82 $\pm$1.01 & 0.15 $\pm$0.04 \\
        &2a& 9.0~~~~~~&1.5$\times$2.4 && 3.53 $\pm$0.17 & 0.79 $\pm$0.14 & 28.76 $\pm$1.41&  3.87 $\pm$0.69 & 0.30 $\pm$0.06 \\
        &3a& 12.6~~~~~~&1.5$\times$4.8 && 3.05 $\pm$0.31 & 0.79 $\pm$0.28 & 16.10 $\pm$1.62&  2.82 $\pm$0.73 & 0.21 $\pm$0.08 \\
        &4a& 16.8~~~~~~&1.5$\times$3.6 && 2.96 $\pm$0.24 & 0.77 $\pm$0.19 & 41.86 $\pm$3.41&  6.71 $\pm$1.13 & 0.21 $\pm$0.09 \\
 
  \hline

MMT$^f$ &5&  8.0~~~~~~&1.5$\times$10.8 && 14.64 $\pm$0.57 & 3.65 $\pm$0.39 & 32.74 $\pm$1.28 & 6.40 $\pm$0.45 & 0.23 $\pm$0.06 \\
        &6&  7.0~~~~~~&1.5$\times$7.2 && 14.32 $\pm$0.47 & 4.00 $\pm$0.50 & 39.75 $\pm$1.32 & 5.97 $\pm$0.48 & 0.16 $\pm$0.07 \\
\hline\hline
\end{tabular}

$^a$distance in arcsec from the brightest H {\sc ii} region. \\
$^b$aperture $x$$\times$$y$ where $x$ is the slit width and $y$ the size 
along the slit in arcsec. \\
$^c$in units 10$^{-16}$\ erg\ s$^{-1}$cm$^{-2}$. \\
$^d$in \AA. \\
$^e$slit orientation with position angle P.A. = --41$^{\circ}$. \\
$^f$slit orientation with position angle P.A. = 0$^{\circ}$. \\
\end{table*}



\begin{table*}[tbh]
\caption{Uncorrected and corrected equivalent widths of the H$\gamma$ and 
H$\delta$ absorption lines in the main body of SBS 0940+544}
\label{t:abshdhg}
\begin{tabular}{lcrcrrrrr} \hline \hline
Telescope &\# of region&Distance$^a$&Aperture$^b$&& 
$EW$(H$\delta$)$^c$  &$EW$(H$\gamma$)$^c$  
&$EW$(H$\delta$)$_{\rm cor}$$^c$  &$EW$(H$\gamma$)$_{\rm cor}$$^c$   \\ \hline

Keck II$^d$ &1& 6.3~~~~~~&1.0$\times$3.0 && 4.51 $\pm$0.22  & 2.57 $\pm$0.23 & 6.41 $\pm$0.49 & 6.37 $\pm$0.55 \\
     &2& 9.1~~~~~~&1.0$\times$2.6 && 6.96 $\pm$0.24 & 5.47 $\pm$0.48 & 7.47 $\pm$0.68 & 6.50 $\pm$0.77 \\
     &3& 12.2~~~~~~&1.0$\times$3.6 && 6.65 $\pm$0.29 & 5.40 $\pm$0.24 & 6.95 $\pm$0.47 & 6.01 $\pm$0.29 \\
     &4& 16.4~~~~~~&1.0$\times$4.8 && 5.71 $\pm$0.53 & 5.40 $\pm$0.41 & 7.19 $\pm$0.79 & 6.40 $\pm$0.41 \\

  \hline

MMT$^d$ &1a& 6.3~~~~~~&1.5$\times$3.0 && 3.95 $\pm$0.50 & 0.70 $\pm$0.60 & 5.99 $\pm$0.78 & 5.08 $\pm$0.68 \\
        &2a& 9.0~~~~~~&1.5$\times$2.4 && 6.28 $\pm$0.43 & 4.50 $\pm$0.47 & 6.80 $\pm$0.59 & 5.61 $\pm$0.55 \\
        &3a& 12.6~~~~~~&1.5$\times$4.8 && 7.63 $\pm$0.63 & 5.23 $\pm$0.70 & 8.00 $\pm$0.69 & 6.00 $\pm$0.74 \\
        &4a& 16.8~~~~~~&1.5$\times$3.6 && 5.76 $\pm$0.87 & 5.01 $\pm$0.52 & 6.73 $\pm$0.92 & 5.81 $\pm$0.68 \\

  \hline

MMT$^e$ &5& 8.0~~~~~~&1.5$\times$10.8 && 5.85 $\pm$0.53 & 2.52 $\pm$0.53 & 6.86 $\pm$0.65 & 4.55 $\pm$0.76 \\
        &6& 7.0~~~~~~&1.5$\times$7.2 && 6.40 $\pm$0.49 & 2.71 $\pm$0.52 & 7.49 $\pm$0.63 & 4.92 $\pm$0.88 \\
\hline\hline
\end{tabular}

$^a$distance in arcsec from the brightest H {\sc ii} region. \\
$^b$aperture $x$$\times$$y$ where $x$ is the slit width and $y$ the size 
along the slit in arcsec. \\
$^c$in \AA. \\
$^d$slit orientation with position angle P.A. = --41$^{\circ}$. \\
$^e$slit orientation with position angle P.A. = 0$^{\circ}$. \\
\end{table*}


\subsubsection{Age from the hydrogen stellar absorption lines 
\label{sec:absorption}}

   Another extinction-insensitive method for determining 
the age of stellar populations is based on equivalent widths of
absorption features. It can be used for larger ages than
the nebular emission line method because the longer-lived B and A stars
show more prominent hydrogen absorption lines.

 Gonzalez Delg\'ado \& Leitherer (\cite{GonLeith99a}) and Gonzalez Delg\'ado,
Leitherer \& Heckman (\cite{GonLeith99b}) calculated a 
grid of synthetic profiles of stellar hydrogen Balmer absorption lines for
effective temperatures and gravities characteristic of starburst galaxies.
They developed evolutionary stellar population synthesis models, synthesizing 
the profiles of the hydrogen Balmer absorption lines from H$\beta$ to H13 
for instantaneous bursts with ages ranging from $10^6$ to $10^9$ yr,  
in the case of a stellar initial mass 
function with a Salpeter slope and mass cutoffs 
$M_{\rm low}$ = 1 $M_{\odot}$ and $M_{\rm up}$ = 80 $M_{\odot}$. Their models 
predict a steady increase of the equivalent widths with age. However, at larger 
ages $\ga$ 1 Gyr the situation is the opposite and the equivalent
widths of the absorption lines decrease with age (Bica \& Alloin 1986).

 The hydrogen absorption lines due to the underlying stellar populations 
are seen clearly along the whole elongated main body of SBS 0940+544.
We show in Fig. \ref{fig:absorb} the H$\gamma$, H$\delta$ and other hydrogen
absorption lines in two regions along the slit oriented with position angle
P.A. = --41$^\circ$. 
A weak K Ca {\sc ii} absorption line is also present. 
While the H$\gamma$ nebular emission line is
superimposed on the absorption line in region 1, it is not seen in region 3.

 We measured the parameters of the H$\gamma$ and H$\delta$ absorption lines
in 4 regions along the main body. 
A careful placement of the continuum is very important for deriving 
accurate $EW$s. To define the continuum level, we select 
several points in the spectral regions free of nebular and stellar lines. The
continuum is then fitted by cubic splines and the quality of the continuum fit 
is visually checked. Although the contamination of the absorption
lines by nebular emission in regions 2, 3 and 4 is very small, not more 
than a few percent, we have corrected the absorption equivalent widths
for it. For this purpose the expected fluxes of the H$\gamma$ and 
H$\delta$ emission lines have been calculated from the flux of the H$\beta$ 
emission line. Then these fluxes have been subtracted from the negative
fluxes of the absorption lines.
Here we adopt theoretical recombination H$\gamma$/H$\beta$ and
H$\delta$/H$\beta$ flux ratios at an electron temperature $T_e$ = $10^4$ K and 
an electron number density $N_e$ = $10^2$ cm$^{-3}$.
We do not correct for extinction. Extinction correction would result in 
lower $EW_{\rm abs}$ and smaller ages. Hence the ages derived in this section
are upper limits.


\begin{figure}[hbtp]
    \psfig{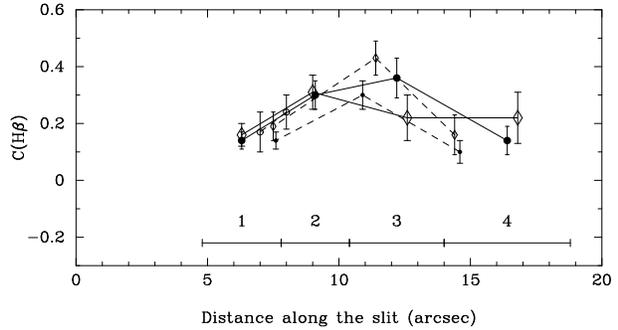}
    \caption{Extinction coefficient $C$(H$\beta$), derived from 
     the flux ratios of the emission lines 
     H$\alpha$ and H$\beta$ in 4 regions along the main body.
     Filled large circles, open large rhombs and solid lines show 
     data respectively from the Keck II and MMT spectra 
     for the slit oriented at
     position angle P.A. = --41$^{\circ}$. Large open circles
     show data from the MMT spectra with the slit oriented at
     position angle P.A. = 0$^{\circ}$.
     Small symbols and dashed lines show the same data, but for 3 regions 
     in larger apertures located at 7\farcs6, 10\farcs9 and 14\farcs6 
from the brightest H {\sc ii} region with
respective apertures 1\farcs0$\times$5\farcs6, 1\farcs0$\times$6\farcs2 and
1\farcs0$\times$8\farcs4 in the Keck II spectra and 
1\farcs5$\times$5\farcs6, 1\farcs5$\times$6\farcs2 and
1\farcs5$\times$8\farcs4 in the MMT spectra.
     }
    \label{fig:chb}
\end{figure}


 We measured equivalent widths for only the H$\gamma$ and H$\delta$ absorption
lines. Although higher-order hydrogen Balmer absorption lines are 
seen in the spectrum of the main body, they are not suitable for 
age determination because of (a) the relatively low signal-to-noise
ratio of the spectra at short
wavelengths and uncertainties in the placement of the continuum in the 
blue region due to many blended absorption features, and 
(b) the weak dependence of 
their equivalent widths on age (Gonzalez Delg\'ado et al. \cite{GonLeith99b}).
Additionally, the H$\epsilon$ absorption line is blended with the H
Ca {\sc ii} absorption line.

 The measured and corrected equivalent widths of H$\gamma$ and H$\delta$ 
absorption lines in the Keck II and MMT spectra are listed in 
Table~\ref{t:abshdhg}. 
The errors include uncertainties in the placement of the continuum.
In Fig.~\ref{fig:hdhg} the measured (dashed lines) 
and corrected (solid lines) equivalent widths of the H$\gamma$ and H$\delta$ 
absorption lines are shown for the 4 regions along the main body of 
SBS 0940+544. 
Note that the corrected equivalent widths of H$\gamma$ and H$\delta$ 
do not show statistically significant spatial variations.

 Fig.~\ref{fig:abshghd} shows the predicted behaviour of the 
equivalent widths of H$\gamma$ (thick solid line) and H$\delta$ 
(thick dashed line) absorption lines with age for 
an instantaneous burst at a metallicity $Z$ = 1/20 $Z_\odot$ 
from Gonzalez Delg\'ado et al. (\cite{GonLeith99b}) for ages $\leq$ 1 Gyr
and Bica \& Alloin (\cite{Bica86}) for ages $>$ 1 Gyr. The upper and lower 
limits of the corrected $EW$(H$\gamma$) and $EW$(H$\delta$) in the 4 regions 
along the main body of the galaxy are shown respectively by thin solid and 
dotted horizontal lines. The shaded rectangular region gives age 
limits, ranging between 15 and 50 Myr. Though slightly larger, this age estimate is
consistent with that obtained from the nebular emission line analysis, 
particularly since it constitutes an upper limit.
Instantaneous burst models also predict low $EW$(H$\gamma$) and
$EW$(H$\delta$) at age $\sim$ 10 Gyr. However, this age is inconsistent 
with the one derived from the emission line equivalent widths.

\subsubsection{Age from the spectral energy distribution \label{SED}}

 Useful constraints on the ages of stellar populations can also be obtained 
from the spectral energy distribution (SED). 
As already pointed out, this method is subject to the 
effects of interstellar extinction. 
However, the combination of the spectral energy distribution method with the
methods discussed in Sect. \ref{sec:emission} and \ref{sec:absorption}
can be used for a simultaneous derivation of the age and extinction.

To fit the observed SEDs, we use the galactic evolution code PEGASE.2 (Fioc \& 
Rocca-Volmerange \cite{F97}) to produce a grid of theoretical SEDs
for an instantaneous burst of star formation and ages ranging between  
0 and 10 Gyr, and a heavy element mass fraction of $Z$ = 1/20 $Z_\odot$.
An initial mass function with a Salpeter 
slope ($\alpha$ = --2.35), and upper and lower mass limits of 120 
$M_\odot$ and 0.1 $M_\odot$ are adopted. 
Because the equivalent widths of hydrogen emission lines in all 4 regions
along the main body are small, the contribution of the ionized gas 
emission is not significant. For this reason we do not include 
gaseous emission in the SED calculations.
Hence, photometric and spectroscopic data give us direct information 
on stellar populations if the interstellar extinction is known.

 The extinction in each region can be estimated from the observed decrement
of the Balmer emission lines. Many strong
hydrogen emission lines are present in the brightest H {\sc ii} region {\it a}
(Fig. \ref{fig:brightsp}) which allow a
precise determination of the extinction. From the Keck II and
MMT spectra of this region we derive very low values of 
$C$(H$\beta$), 0.08 $\pm$ 0.02 and 0.00 $\pm$ 0.01 respectively.
The H$\alpha$ and H$\beta$ emission lines
are also present in the main body, though weaker. We use the fluxes of these lines to derive
$C$(H$\beta$) in different regions along the slit. 
The results are shown in Table \ref{t:emhahb} and in Fig. \ref{fig:chb}. We 
note that, in general, the extinction in the main body, derived from the 
observed Balmer decrement, is significantly larger than that in the brightest
H {\sc ii} region, with a maximum $C$(H$\beta$) $\sim$ 0.3 
in region 3 at a distance of $\sim$ 12\arcsec\ (or $\sim$ 1.5 kpc) 
from the brightest H {\sc ii} region.

\begin{figure}[hbtp]
    \psfig{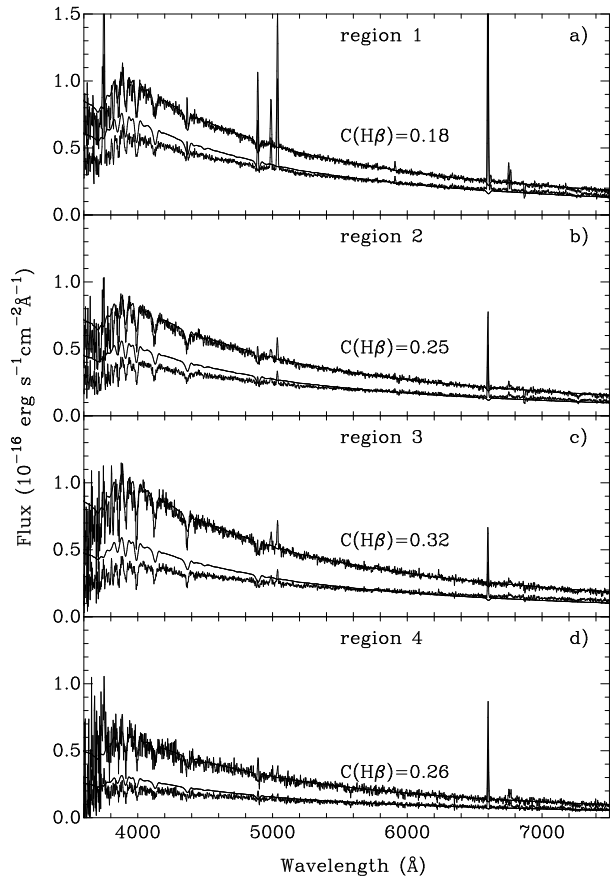}
\caption{Keck II telescope spectra of 4 regions in the main body of 
SBS 0940+544 on which synthetic continuum spectral energy distributions are 
superposed. Lower spectra in (a) -- (d) are synthetic SEDs of a 20
Myr instantaneous burst stellar population superposed on the observed spectra 
uncorrected for extinction ($C$(H$\beta$) = 0).
Upper spectra in (a) -- (d) are synthetic SEDs of a 20 Myr 
instantaneous burst stellar
population superposed on the spectra corrected for extinction.
Each upper spectrum is labeled by the extinction coefficient $C$(H$\beta$)
which gives the best fit of the theoretical SED to the extinction-corrected
observed SED.
     }
    \label{fig:spfit}
\end{figure}

In principle, the extinction can also be derived from the SEDs
of regions 1 -- 4, adopting the age of the stellar population derived
from the equivalent widths of the hydrogen emission and absorption lines.
First we assume $C$(H$\beta$) = 0.
Then the observed SEDs are redder than 
the theoretical SED of a stellar population with age $t$ = 20 Myr 
(lower spectra in Fig. \ref{fig:spfit}a -- \ref{fig:spfit}d). 
   One of the reasons for the difference between the observed and theoretical
SEDs is that 
interstellar extinction in the main body is not negligible and it
modifies the observed spectral energy distribution. 
To derive $C$(H$\beta$) we adopt an age
of the stellar population in each of the regions 1 -- 4 equal to 20 Myr,
as a representative value derived from the hydrogen 
emission and absorption line equivalent
widths. We then compute the extinction coefficient $C$(H$\beta$) to achieve the
best agreement between the observed SED after correction for interstellar 
extinction and the theoretical SED. The observed extinction-corrected
SEDs are superimposed on the synthetic
20 Myr stellar population SED for regions 1 to 4 in Fig. 
\ref{fig:spfit}a -- \ref{fig:spfit}d (upper spectra). They are labeled by
the derived values of $C$(H$\beta$). These values are consistent 
with those derived from the H$\alpha$/H$\beta$ flux ratios
(Table \ref{t:emhahb}, Fig. \ref{fig:chb}). 
However, despite this consistency, it is seen from Fig. 
\ref{fig:spfit}, that the observed SEDs
(upper spectra) at $\lambda$ $\la$ 3900\AA\ are not 
well fitted by the theoretical SEDs. Hence, we conclude that the instantaneous 
burst models do not adequately reproduce the observed properties of 
the SBS 0940+544 main body.

\subsection{The case of continuous star formation\label{extend}}

Our estimates for the stellar population age in SBS 0940+544 made in Sect. 
\ref{inst} are based on the assumption of 
an instantaneous burst of star formation. 
Now we consider how the age of the 
stellar population is changed if star formation is continuous. 
For this we adopt a constant star formation rate in the interval between the 
time $t_{\rm i}$ when star formation starts and $t_{\rm f}$ when it stops.
Time is zero now and increases to the past. 
It is customary to consider models with a constant star formation
rate, with star formation continuing until now, i.e. with $t_{\rm f}$ = 0.
However, we will also consider models with $t_{\rm f}$ $>$ 0 because of the 
following reason. The number of the ionizing photons in regions 1 -- 4
derived from the H$\alpha$ luminosity is in the range 
$N$(Lyc) = (2 -- 7) $\times$ 10$^{49}$ s$^{-1}$, several times lower than
$N$(Lyc) emitted by a single star with mass $\sim$ 100 $M_\odot$ 
(Vacca et al. \cite{Vacca96}). Hence, an IMF with a high $M_{\rm up}$ 
does not strictly apply. There are however several
circumstances which may explain the lack of ionizing photons: 
(a) stochastic effects on the IMF; (b) the major part
(more than 50\%) of the ionizing photons escapes the H {\sc ii} region or
is absorbed by dust; (c) the IMF upper mass limit is truncated  
at $\sim$ 30 $M_\odot$; (d) star formation in regions 1 -- 4
stopped several Myr ago, i.e. $t_{\rm f}$ $>$ 0.
We have considered all of these possibilities (except for (a)) in fitting the
data and found that the result is not changed significantly. 
Therefore, in the subsequent discussion 
we use models with $M_{\rm up}$ = 120$M_\odot$ and 
$t_{\rm f}$ $\geq$ 0.


We use the model equivalent widths of hydrogen emission 
and absorption lines and SEDs for instantaneous bursts (Fioc \& 
Rocca-Volmerange \cite{F97}; Gonzalez Delg\'ado et al. \cite{GonLeith99b};
Bica \& Alloin \cite{Bica86})
to calculate the temporal evolution of the equivalent widths of hydrogen 
emission and absorption lines in the case of continuous constant star 
formation. 
The results are given in Fig. \ref{fig:ewhbha} and \ref{fig:abshghd}. 
The temporal dependences of the equivalent
widths of the H$\beta$ and H$\alpha$ emission lines (Fig. \ref{fig:ewhbha}), 
and the H$\delta$ and H$\gamma$ absorption lines (Fig. \ref{fig:abshghd}) 
are shown for continuous star
formation starting at time $t_{\rm i}$, as defined by the abscissa value, and 
stopping at $t_{\rm f}$ = 0 Myr (dotted line) and $t_{\rm f}$ = 8 Myr
(dash-dotted line). 
The equivalent widths of the H$\beta$ and H$\alpha$ emission
lines and of the H$\delta$ and H$\gamma$ absorption lines 
in the spectrum of the stellar population formed between $t_{\rm i}$ and 
$t_{\rm f}$
are represented in Fig. \ref{fig:ewhbha} and \ref{fig:abshghd} by the values 
of $EW$ at time $t_{\rm i}$. At a fixed $EW$, it is seen that the younger the 
youngest stars, the larger the time interval $t_{\rm i} - t_{\rm f}$, 
i.e. the older the oldest stars.

\subsubsection{Continuous star formation with young stellar population 
\label{young}}

\begin{figure}[hbtp]
    \psfig{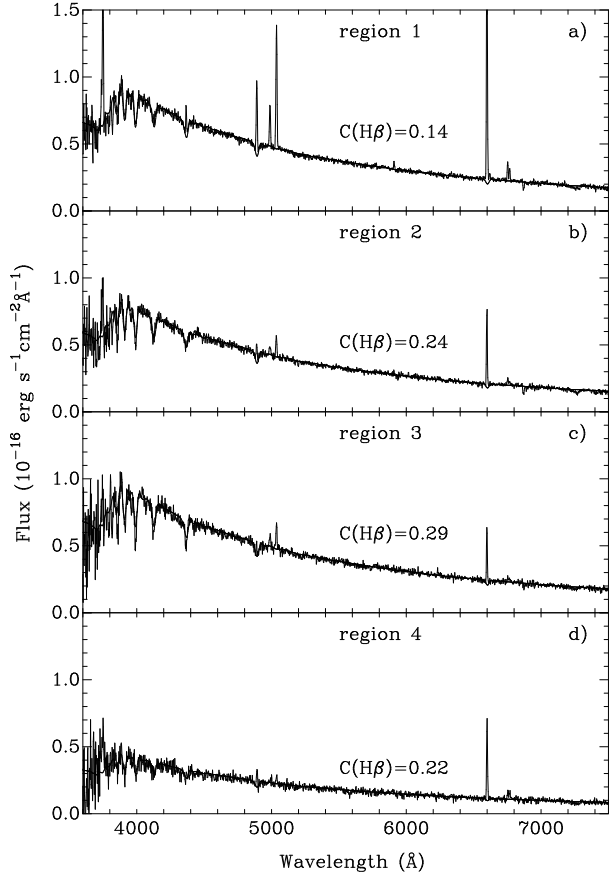}
\caption{Keck II telescope spectra of 4 regions along the main body of 
SBS 0940+544 on which synthetic continuum spectral energy distributions are 
superposed. Spectra in (a) -- (d) are synthetic SEDs of a stellar
population continuously formed with constant star formation rate
between 8 and 100 Myr ago. These SEDs are superposed on the spectra 
corrected for extinction.
Each spectrum is labeled by the extinction coefficient $C$(H$\beta$)
which gives the best fit of the theoretical SED to the extinction-corrected
observed SED.
}
    \label{fig:spfitext}
\end{figure}

Can the observed SEDs of regions 1 -- 4 be fitted with only a young stellar
population continuously formed over the last 100 Myr?
To fit the observed SEDs and derive the extinction in regions 1 -- 4
of the main body in the case of continuous star formation, we consider 
star formation occurring between $t_{\rm f}$ = 8 Myr and $t_{\rm i}$ = 100 
Myr. This model predicts $EW$(H$\delta$) = 7.2\AA, $EW$(H$\gamma$) = 6.0\AA, 
and $EW$(H$\beta$) = 3.5\AA, $EW$(H$\alpha$) = 21.2\AA, 
close to the values observed in regions 2 and 3 (Tables \ref{t:emhahb} and
\ref{t:abshdhg}). However, the observed emission line equivalent
widths in regions 1 and 4 are larger than the predicted
values. For these two regions, a more appropriate model is that with $t_{\rm i}$
$<$ 100 Myr and/or $t_{\rm f}$ $<$ 8 Myr. 
We show in Fig. \ref{fig:spfitext}
the results of our fitting. As in the case of an instantaneous burst,
we adjust the extinction coefficient $C$(H$\beta$) to achieve the
best agreement between the observed SED after correction for interstellar 
extinction and the theoretical SED. The observed extinction-corrected
SEDs are superimposed on the synthetic
stellar population SED for regions 1 to 4 in Fig. 
\ref{fig:spfitext}a -- \ref{fig:spfitext}d. They are labeled by
the derived values of $C$(H$\beta$). These values 
are in agreement with those derived from the H$\alpha$/H$\beta$ flux ratio.
Note that the synthetic SEDs with continuous star formation 
reproduce better the observed spectra in the spectral 
range $\lambda$ $<$ 3900\AA\ than those calculated on the assumption of an
instantaneous burst.
This makes continuous star formation in the main body of SBS 0940+544, 
occuring during the last 100 Myr, a possible interpretation.
%


\begin{figure}[hbtp]
    \psfig{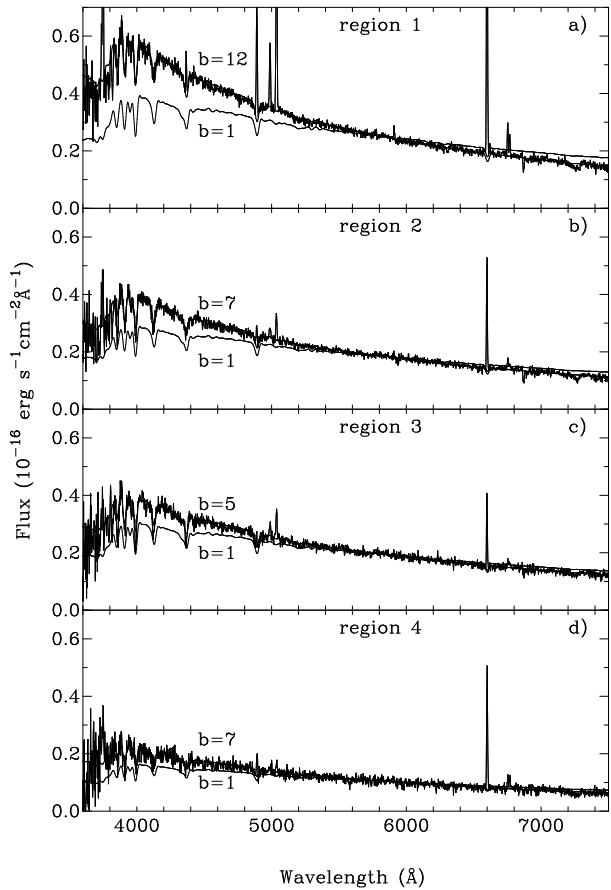}
\caption{Keck II telescope spectra of 4 regions in the main body of 
SBS 0940+544 on which synthetic continuum 
spectral energy distributions are superposed. 
Synthetic SEDs shown in panels (a) -- (d) and labeled by $b$ $\equiv$ 
SFR($t$ $\leq$ 100 Myr)/SFR($t$ $>$ 100 Myr) = 1 
correspond to stellar populations formed continuously with a constant star 
formation rate between 0 and 10 Gyr. Synthetic spectra labeled by $b$ $>$ 1
correspond to stellar populations formed continuously between 0 and 10 Gyr
with a star formation rate enhanced by a factor of $b$ during the last 100 Myr.
These SEDs are superposed on the spectra uncorrected for extinction.
     }
    \label{fig:spfitchb0}
\end{figure}


\begin{figure}[hbtp]
    \psfig{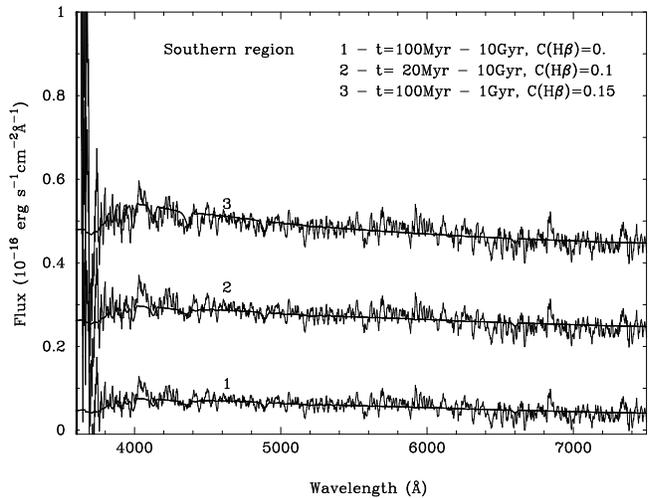}
    \caption{MMT spectrum of region {\it d} in the LSB host of SBS 0940+544,
smoothed by a 5-point box-car and labeled as ``1''. The spectrum 
is extracted from the two-dimensional spectrum obtained with the slit
oriented at position angle P.A. = 0$^\circ$ in an aperture 
4\farcs8 $\times$ 1\farcs5.
A theoretical SED of a stellar population continuously formed between
100 Myr and 10 Gyr ago is superposed on ``1''.
The spectra labeled as ``2'' and ``3'' are the same spectra corrected for
extinction respectively with $C$(H$\beta$) = 0.1 and 0.15 and superposed on a 
theoretical SEDs of a stellar population continuously formed between
100 Myr and 1 Gyr ago and 20 Myr and 1 Gyr ago. The latter two spectra and 
SEDs are offset up by 2$\times$10$^{-17}$ erg s$^{-1}$cm$^{-2}$\AA$^{-1}$
and 4$\times$10$^{-17}$ erg s$^{-1}$cm$^{-2}$\AA$^{-1}$.
     }
    \label{fig:spold}
\end{figure}


\subsubsection{Continuous star formation including old stellar population 
\label{chb0}}

We consider next different continuous star formation scenarios in which 
an old stellar population is present in the main body of SBS 0940+544.
For this, we again adopt $C$(H$\beta$) = 0 and consider models with constant and varying 
SFRs with ages ranging between 0 and 10 Gyr. 
Specifically, for a varying SFR, we consider two periods of continuous
star formation with constant but different SFRs, occurring in the age interval 
$\leq$ 100 Myr and $>$ 100 Myr.
To quantify the recent-to-past star formation rate ratio, we introduce the
  parameter $b$ = 
SFR($t$ $\leq$ 100 Myr)/SFR($t$ $>$ 100 Myr). First we consider models with
constant star formation during the whole 0 -- 10 Gyr range, i.e. models with $b$ = 1.
In Fig. \ref{fig:spfitchb0}a -- \ref{fig:spfitchb0}d we show such SEDs 
superimposed on the observed spectra of regions 1 -- 4. It is evident
that these models do not reproduce the observed SEDs. However, by
increasing the parameter $b$ we can fit the observed SEDs. 
These theoretical SEDs are labeled by $b$ $>$ 1 
in Fig. \ref{fig:spfitchb0}a -- \ref{fig:spfitchb0}d. The predicted
equivalent widths of the emission and absorption hydrogen lines are also 
in agreement with the observed ones. If non-negligible extinction is present
in regions 1 -- 4 then to fit the observed SEDs, the parameter $b$ 
should be further increased. 
In particular, if the extinction $C$(H$\beta$) =
0.28 derived for region 2 from the H$\alpha$/H$\beta$ flux ratio is adopted then
the parameter $b$ should be as high as $\sim$ 100 to fit observations.

In summary, analysis of the spectroscopic data for the main body of
SBS 0940+544 shows that 
the stellar population can be equally well reproduced by a model
of continuous star formation during the last 100 Myr, or by a model in
which stars are continuously formed in the period 0 -- 10 Gyr.
In the former case, the extinction, as derived from the 
H$\alpha$/H$\beta$ flux ratio, should be taken into account. 
In the latter case, a fit to the observed SED is only possible 
when $b>$5, i.e. the star formation 
rate in the main body of SBS 0940+544 has significantly increased
over the last 100 Myr. 

\subsubsection{Age of the low-surface-brightness host
  \label{old}}

An upper limit to the age of the stellar population in SBS 0940+544
can in principle be derived from photometric and spectroscopic measurements
of the faint ($\mu_V\approx 24 - 24.5 $ mag arcsec$^{-2}$) region {\it d} (cf. Fig.~\ref{f1}a).

As discussed in Sect. \ref{morph} the $V-I$ and $V-R$ colours in this region are
only slightly redder than the average respective values of 0.58 mag and 0.33 mag derived
for the outskirts of SBS 0940+544 from Fig. \ref{f4}, thus it may be 
considered representative for the stellar LSB component. 

The spectrum of region {\it d} extracted from the MMT \#2 spectrum
(position angle P.A. = 0$^\circ$)
(cf. Fig. \ref{f1}b) within a 4\farcs8 $\times$ 1\farcs5 aperture
is shown in Fig. \ref{fig:spold} and labeled ``1''.
Because of its very low intensity, the spectrum is smoothed by a
5-point box-car. No appreciable emission or absorption lines are seen
in the spectrum, probably because of the low signal-to-noise ratio. Therefore, 
we cannot estimate an age of the stellar populations in region {\it d} by 
extinction-independent methods, or decide on whether the red 
colours of this region are caused by enhanced extinction or by an 
intrinsically red stellar population. 
Spectrum ``1'', when not corrected for extinction,
is fitted by a theoretical SED of a stellar population
continuously formed with constant star formation rate between 100 Myr and
10 Gyr ago. 

However, in the presence of extinction the range of ages can be 
increased
or the upper age limit can 
be significantly reduced. If a value of $C$(H$\beta$) = 0.1 
is assumed for region {\it d} then the corrected spectrum (labeled
``2'' in Fig. \ref{fig:spold}) is fitted by a theoretical SED of a 
stellar population continuously formed with constant star formation rate 
between 20 Myr and 10 Gyr ago. However, it is unlikely to reduce the lower
age limit to values less than 10 -- 15 Myr, otherwise emission lines must be 
present in the spectrum of region {\it d}. 
If a value of $C$(H$\beta$) = 0.15 
is assumed then the corrected spectrum (labeled
``3'' in Fig. \ref{fig:spold}) is fitted by the theoretical SED of a 
stellar population continuously formed with a constant star formation rate 
between 100 Myr and 1 Gyr ago.
The poor signal-to-noise ratio of the spectrum
of region {\it d} precludes reliable estimates of the age for this region.
 
\subsubsection{Age from the surface brightness and colour distributions 
\label{coldist}}

As emphasized in the previous discussion,
photometric data alone do not allow to draw definite conclusions on
the age of the stellar populations. However, they can provide a consistency check of the stellar 
population ages derived from the spectra. Are the ages of the stellar 
populations derived above compatible with the broad-band colours?

\begin{figure*}[hbtp]
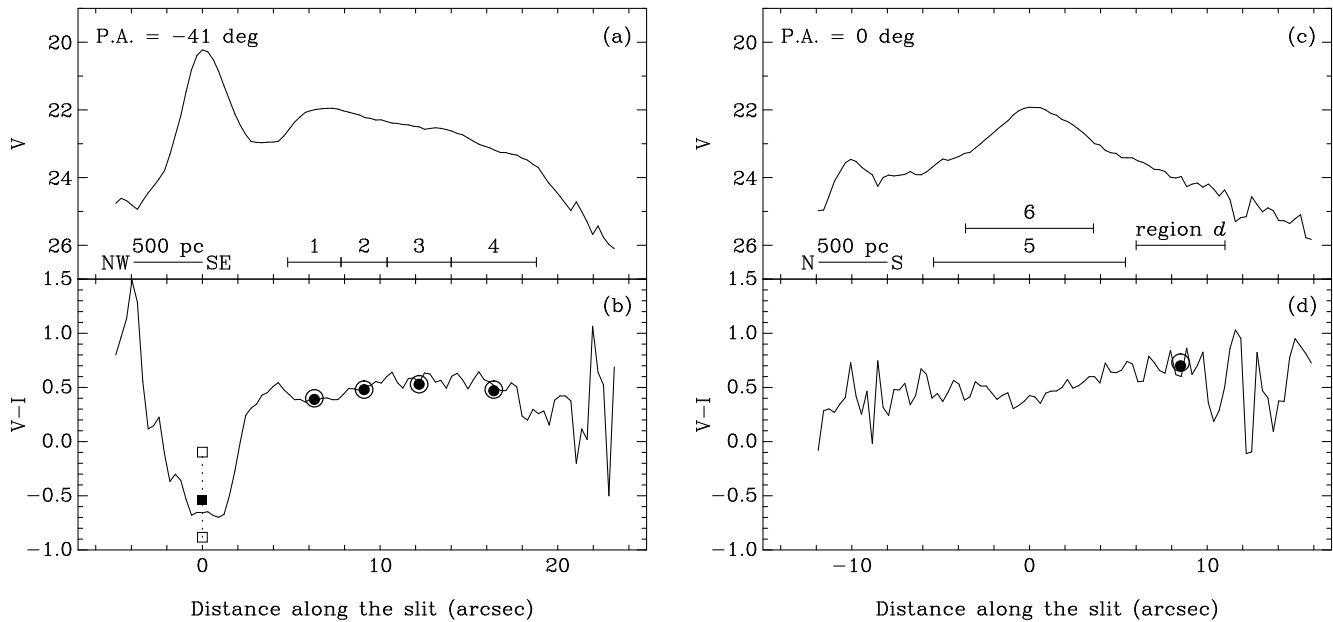

  \hspace*{0.1cm}\psfig{figure=MS1281f15a.eps,angle=0,width=8.5cm}
  \hspace*{0.5cm}\psfig{figure=MS1281f15b.eps,angle=0,width=8.5cm}
    \caption{ (a) The $V$ surface brightness distribution 
along the slit oriented at position angle P.A. = --41$^\circ$. The origin is 
at the location of the brightest H {\sc ii} region. The regions used for
spectroscopic analysis are labeled in (a) ``1'' to ``4''.
(b) The $V-I$ colour distribution along the slit with position angle
P.A. = --41$^\circ$. Filled circles are the predicted colours of a
stellar population continuously formed with a constant star formation rate
between 8 Myr and 100 Myr ago and reddened with an extinction coefficient as 
derived for each region in Fig. \ref{fig:spfitext}, 
open circles are the predicted colours of the stellar populations
continuously formed between 0 and 10 Gyr with the enhancement parameters
$b$ as derived for each region in Fig. \ref{fig:spfitchb0}.
The filled square indicates the colour of a 4 Myr stellar 
population to which has 
been added the observed ionized gas emission in the brightest H {\sc ii} 
region. 
The upper open square shows the colour of a pure 4 Myr stellar population, 
while the lower open square shows the colour of pure ionized gas emission. 
(c) The $V$ surface brightness distribution along the
slit oriented at position angle P.A. = 0$^\circ$. The origin is 
at the location of the surface brightness maximum in the main body
coincident with region {\it c} in Fig.~\ref{f1}.
The regions studied spectroscopically
are labeled as ``5'' and ``6''. The location of the reddest part 
of SBS 0940+544 is marked as ``region {\it d}''.
(d) The $V-I$ colour distribution along the slit with position angle
P.A. = 0$^\circ$. The filled circle shows the predicted colour of a
stellar population continuously formed with constant star formation rate
and reddened with an extinction coefficient $C$(H$\beta$) = 
0.15 as derived for the reddest region in Fig. \ref{fig:spold} (spectrum 3)
while the open circle shows the predicted colour of a
stellar population continuously formed with constant star formation rate
and reddened with an extinction coefficient $C$(H$\beta$) = 
0.1 as derived for the reddest region in Fig. \ref{fig:spold} (spectrum 2).
 }
    \label{fig:colors}
\end{figure*}

We derived $V$ and $I$ surface brightness and colour distributions  for
the regions covered by the spectroscopic observations at both slit 
orientations and compared them with predictions from our
population synthesis modeling. The results of this comparison are shown in
Fig. \ref{fig:colors}. The predicted colours
obtained from convolving the theoretical SEDs
with the appropriate filter bandpasses are shown by different symbols.
The transmission curves for the Johnson $V$ and Cousins $I$ bands are taken
from Bessel (\cite{B90}). The zero points are from Bessel, Castelli \& Plez
(\cite{B98}). 

Since the contribution of the ionized gas emission to the total brightness of 
the brightest H {\sc ii} region ({\it a}) of SBS 0940+544 is high, the 
theoretical SED for this region has been constructed in the following way. We 
have used the 4 Myr stellar population SED for a heavy element mass fraction
$Z$ = $Z_\odot$/20. The {\it observed} gaseous SED
is then added to the calculated stellar SED, its 
contribution being determined by the ratio of the observed equivalent width
of the H$\beta$ emission line to the one expected for pure gaseous emission.
To calculate the gaseous continuum SED, the observed
H$\beta$ flux and the electron temperature have been derived from the optical
spectrum (Tables \ref{t:Intens} and \ref{t:Chem}). The contribution of the
free-bound, free-free and two-photon continuum emission has been taken into
account for the spectral range from 0 to 5 $\mu$m (Aller \cite{Aller84}; 
Ferland \cite{F80}). Emission lines are superposed on the gaseous continuum
SED with intensities derived from spectra in the spectral range 
$\lambda$3700 -- 7500 \AA. Outside this range, the intensities of emission 
lines (mainly hydrogen lines) have been calculated from the 
extinction-corrected flux of H$\beta$.


As evident from Fig. \ref{fig:colors}b the observed colour 
of the brightest H {\sc ii} region
is very blue $V-I$ $\sim$ --0.6 mag and cannot be reproduced by a stellar
population of any age. 
In particular, the $V-I$ colour equal to --0.1 mag of a 4 Myr old stellar population
(upper open square) is significantly redder than that observed.
However, the synthetic colour of a 4 Myr stellar population together with the
ionized gas (filled square) is very similar to the observed colour. We also
show by the lower open square the colour of pure ionized gas emission.
It is very blue ($V-I$ $\sim$ --0.9 mag) if the observed intensities of 
the emission lines (Table \ref{t:Intens}) and the redshift of SBS 0940+544 are 
taken into account.

   We already pointed out that the contribution of the gaseous emission is 
small in the main body. Therefore, we do not take it into account and consider
the colours of a stellar population continuously formed with a constant star
formation rate between 8 Myr and 100 Myr ago and reddened by interstellar 
extinction. In Fig. \ref{fig:colors}b we show by filled circles the
colours of four regions with extinction coefficients derived from the best
of the theoretical fits to the observed extinction-corrected spectra 
(spectra in Fig. \ref{fig:spfitext}a - \ref{fig:spfitext}d).
Open circles show the predicted 
colours of a stellar population continuously formed between 0 and 10 Gyr ago,
assuming enhanced star formation during last 100 Myr, as defined by
parameter $b$ in Fig. \ref{fig:spfitchb0}a - \ref{fig:spfitchb0}d (upper 
spectra). The agreement between the $V-I$ colours
obtained from the photometric data and those derived from the spectral energy
distributions in both scenarios is very good.

We also compare the colour of the southern region {\it d} 
(Fig. \ref{fig:colors}d) with predictions. Because of the noisy spectrum
and large photometric errors for this region a broad range of ages can
be consistent with the observed colour. In particular, the colour of the 
stellar population continuously formed with a constant star formation rate 
between 100 Myr and 1 Gyr ago and reddened with extinction $C$(H$\beta$) = 0.15
is shown in Fig. \ref{fig:colors}d by the filled circle, while the colour 
of the stellar population continuously formed with a constant star formation 
rate between 20 Myr and 10 Gyr ago and reddened with extinction 
$C$(H$\beta$) = 0.1 is shown by the open circle. They are in good
agreement with the observed colour for this region.

   Finally, an upper age limit for stellar populations in SBS 0940+544
can be estimated from the $V-R$ and $V-I$ colours of the outermost regions
without spectroscopic observations. 
The mean colours in the outskirts of the LSB component 
of $V-R$ = 0.33 $\pm$ 0.04 and $V-I$ = 0.58 $\pm$ 0.03 (Fig. \ref{f4}) are
compatible with those of a stellar population forming continuously for the last several Gyr
and assuming that star formation continues until now.
However, the spatial distributions of the old and young stellar 
populations inferred from HST colour-magnitude 
diagrams of some galaxies, are different 
(e.g. Schulte-Ladbeck et al. \cite{SCH98}). The
oldest stars are seen at largest distances, where younger stars are not
detected. It is likely that stars in these extended haloes were formed during
the galaxy formation era and no new stars have formed since then, or stars have
diffused to haloes from the inner regions where they were born. If this is the
case for the outermost regions of SBS 0940+544, then to explain their observed
$V-R$ and $V-I$ colours, the upper age limit should be significantly reduced,
because no young stellar population is present there. 
In that case, the observed colours would be compatible 
with an upper age limit of $\leq$ 1 Gyr for the stellar LSB component.

\section{Conclusions}

  We present in this paper a detailed photometric and spectroscopic study 
of the very metal-deficient blue compact dwarf galaxy SBS 0940+544 
($Z$ $\sim$ $Z_\odot$/27), a good young galaxy candidate. Photometric $V$ and $I$
data have been obtained with the 2.1m Kitt Peak telescope, and $R$
imaging has been done with the 1.23m Calar Alto telescope. Very high 
signal-to-noise spectra in the optical range have been obtained with the 
10m Keck II telescope and the 4.5m MMT at two slit orientations. 
We have reached the following conclusions:

\begin{enumerate}

\item SBS 0940+544 is a nearby BCD galaxy
with {\it cometary}-like structure, i.e. a bright off-center H {\sc ii} region
and an elongated main body. The $V$, $R$ and $I$ surface brightness profiles 
of the galaxy's main body are very similar, with an 
exponential scale length $\alpha$ $\sim$ 320 pc. The $V-I$ colour of the 
brightest H {\sc ii} region is very blue
$\sim$ --0.6 mag, due to the combined effect of the young stellar population
and the ionized gas emission. The colours of the main body are much redder,
without significant gradients.

\item The derived oxygen abundance 12 + log(O/H) = 7.46 -- 7.50 is consistent
within the errors with previous abundance determinations.
The $\alpha$-element Ne/O, S/O, Ar/O abundance ratios are 
in good agreement with the mean ratios derived from previous studies of
BCDs (Izotov \& Thuan \cite{IT99}).
The nitrogen-to-oxygen abundance ratio log N/O = --1.60 - --1.65 
lies in the narrow range obtained
by Thuan et al. (\cite{til95}) and Izotov \& Thuan (\cite{IT99}) for the most
metal-deficient BCDs. These abundances suggest that SBS 0940+544 is a good
local young galaxy candidate. 

\item The $^4$He mass fractions $Y$ = 0.247 $\pm$ 0.003 and 0.244 $\pm$ 0.007
derived respectively from the Keck II and MMT observations of the
brightest H {\sc ii} region in SBS 0940+544 are in good agreement with previous
determinations of the helium abundance in this galaxy and are consistent with
the value of the primordial $^4$He abundance $Y_p$ = 0.244 -- 0.245
derived by Izotov \& Thuan (\cite{IT98a}) and Izotov et al. (\cite{ICFGGT99}).

\item The hydrogen H$\alpha$ and H$\beta$ lines in the main body are seen 
in emission, while higher-order Balmer lines are seen in absorption.
Three methods are used to constrain the age
of the stellar populations in the main body of SBS 0940+544. The first one
is based on the age dependence of the H$\alpha$ and H$\beta$ emission line
equivalent widths, the second one on the age
dependence of the hydrogen H$\gamma$ and H$\delta$ absorption line equivalent
widths, and the third one on the age dependence of the spectral energy
distribution. Several star formation histories have been considered.
The single instantaneous burst models do not reproduce the observed
SEDs, suggesting that star formation in the main body was continuous.
We find that models of continuous star formation with a constant star
formation rate starting 10 Gyr ago are also excluded. However, models 
with continuous star formation during the period 0 -- 10 Gyr, and with a varying star
formation rate are able to explain the observed properties. In particular,
models in which the star formation rate during the last 100 Myr has increased
several times, can reproduce both observed equivalent widths of the emission and
absorption hydrogen lines and SEDs. However, the observed spectroscopic 
and photometric characteristics are reproduced equally well by models in which
stars were continuously formed during the last 100 Myr only, 
if an extinction correction as 
derived from the H$\alpha$/H$\beta$ flux ratio is applied to the SEDs.

\item 
The age of the reddest very low-intensity southern region in 
the SBS 0940+544 main body gives an upper limit to the 
galaxy's age. 
Because of the region's faintness, we were not able to detect 
hydrogen emission and/or absorption features to derive a reliable age.
Assuming no extinction, we find that the SED can be fitted by a stellar population
continuously formed with a constant star formation rate for a whole Hubble
time. 
However, some extinction is likely to exist. In particular, if an
extinction coefficient $C$(H$\beta$) = 0.15 is assumed,
the extinction-corrected spectral energy distribution of the southern region
can be well fitted by a model with a stellar population continuously formed 
with a constant star formation rate between 100 Myr and 1 Gyr ago.

In summary, we find no compelling evidence which favours either a young
or an old age for SBS 0940+544.

\end{enumerate}

\begin{acknowledgements}
We thank 
referee G\"oran \"Ostlin for valuable comments and criticism on the 
manuscript. N.G.G. has been supported by DFG grant 
436 UKR 17/1/00 and Y.I.I. acknowledges the G\"ottingen Academy of Sciences
for a Gauss professorship.
N.G.G. and Y.I.I. have been partially supported by INTAS 97-0033 and 
Swiss SCOPE 7UKPJ62178 grants. They are grateful for the hospitality 
of the G\"ottingen Observatory. 
Those authors, P.P. and K.J.F acknowledge support by the Volkswagen 
Foundation under grant No. I/72919. Y.I.I. and T.X.T have been partially 
supported by NSF grant AST-9616863. Research by P.P. and K.J.F. has been supported by the
Deutsches Zentrum f\"{u}r Luft-- und Raumfahrt e.V. (DLR) under
grant 50\ OR\  9907\ 7. C.B.F. acknowledges the support of the NSF under grant
AST-9803072, and K.G.N. of the Deutsche Forschungsgemeinschaft (DFG) grant FR325/50-1.
\end{acknowledgements}

\end{document}